\documentclass[12pt,onecolumn,draftclsnofoot]{IEEEtran}
\usepackage[utf8]{inputenc}
\usepackage{textcomp}
\usepackage{xspace}
\usepackage{graphicx}
\usepackage{amsmath}
\usepackage[version=4]{mhchem}
\usepackage{siunitx}
\usepackage{subcaption}
\usepackage{longtable,tabularx}
\usepackage{placeins}
\usepackage{cuted}
\usepackage{booktabs}

\usepackage{hyperref}
\ifpdf
\hypersetup{
  colorlinks=true,
  linkcolor=blue,
  citecolor=blue,
  urlcolor=blue,
  pdftitle={Square-Root Higher-Order Unscented Estimators for Robust Orbit Determination},
  pdfauthor={} 
}
\fi
\usepackage{algorithm2e}
\usepackage{adjustbox}
\usepackage{multirow}
\usepackage{ragged2e}
\usepackage{setspace}
\ifCLASSINFOpdf
\else
\fi

\begin{document}
%
\title{Square-Root Higher-Order Unscented Estimators for Robust Orbit Determination}
%
%
%

\author{Yang~Yang,~\IEEEmembership{Senior Member,~IEEE,}
\thanks{Dr Y. Yang is with the School of Mechanical and Manufacturing Engineering,
University of New South Wales, Sydney
, NSW 2052, Australia. E-mail: yiyinfeixiong@gmail.com.}
}

%
%

\markboth{Manuscript for IEEE Transactions on Aerospace and Electronic Systems}%
{Shell \MakeLowercase{\textit{et al.}}: Bare Demo of IEEEtran.cls for IEEE Journals}
%



\maketitle

\begin{abstract}
    Orbit determination (OD) is a fundamental problem in space surveillance and tracking, crucial for ensuring the safety of space assets. Real-world ground-based optical tracking scenarios often involve challenges such as limited measurement time, short visible arcs, and the presence of outliers, leading to sparse and non-Gaussian observational data. Additionally, the highly perturbative and nonlinear orbit dynamics of resident space objects (RSOs) in low Earth orbit (LEO) add further complexity to the OD problem.
    This paper introduces a variant of the higher-order unscented Kalman estimator (HOUSE) called $w$-HOUSE, which employs a square-root formulation and addresses the challenges posed by nonlinear and non-Gaussian OD problems. The effectiveness of $w$-HOUSE was demonstrated through synthetic and real-world measurements, specifically outlier-contaminated angle-only measurements collected for the Sentinel 6A satellite flying in LEO. Comparative analyses are conducted with the original HOUSE (referred to as $\delta$-HOUSE), unscented Kalman filters (UKF), conjugate unscented transformation (CUT) filters, and precise orbit determination solutions estimated via onboard global navigation satellite systems measurements. The results reveal that the proposed $w$-HOUSE filter exhibits greater robustness when dealing with varying values of the dependent parameter compared to the original $\delta$-HOUSE. Moreover, it surpasses all other filters in terms of positioning accuracy, achieving three-dimensional root-mean-square errors of less than \SI{60}{\meter} in a three-day scenario. This research suggests that the $w$-HOUSE filter represents a viable alternative to UKF and CUT filters, offering improved positioning performance in handling the nonlinear and non-Gaussian OD problem associated with LEO RSOs.
\end{abstract}


\section{Introduction}
\label{sec:introduction}

The escalating presence of space debris in Earth's orbit poses an ever-growing threat to the safety of space assets. In order to mitigate the risk of collisions with these hazardous objects, it is imperative to obtain highly accurate and reliable orbital information about them. Given the proliferation of resident space objects (RSOs), ground-based optical tracking has emerged as a crucial component of space surveillance. However, the operation of traditional optical sensors is contingent on specific visibility conditions, where RSOs are sunlit while the background is dark. Consequently, these facilities primarily operate during nighttime hours. 
Moreover, various practical challenges complicate the process of gathering orbital data through ground-based optical tracking. These challenges include constraints on the field of view of telescopes, the rapid motion of RSOs, particularly those in low Earth orbit (LEO), and the limited scheduling opportunities for tracking each object. As a result, the number of visible arcs, or tracks of measurements, is often limited, and each track contains only short intervals of effective measurements, often spanning just a few minutes. The complexity of orbit determination (OD) is further amplified when there is a substantial data gap, such as a day or more, between consecutive tracks of measurements. Additionally, optical sensor measurements are typically subject to various types of noises, including background noise, shot noise, readout noise, dark current noise, and quantisation noise~\cite{tan2021quantifying}, which can introduce non-Gaussian characteristics into angular measurements like right ascension and declination. Furthermore, the imperfect nonlinear orbital dynamics degrades the prediction accuracy and contributes to filter divergence. 

The combination of these challenges transforms the task of OD for LEO RSOs into a formidable problem, characterised by nonlinearity and non-Gaussianity. To address such complex estimation problems, various filtering techniques have been developed over the years. 
The historical roots of online, real-time state estimation based on dynamic models can be traced back to Rudolf E. Kalman's pioneering work in 1960~\cite{kalman1960new}, where he introduced the Kalman filter (KF) for linear stochastic systems with Gaussian assumptions. This enabled first publicly known application in the Apollo lunar mission for the estimation of the position of the spacecraft~\cite{schmidt1981kalman}. Since then, numerous KF variants have been devised and widely employed in space navigation and OD problems, some capable of handling nonlinear and non-Gaussian estimation scenarios. 

Nonlinearities in the system dynamics can be handled by approximating them via first-order Taylor expansions~\cite{welch1995introduction, lam2012analysis}, leading to the development of extended Kalman filters (EKFs). However EKFs may fail when the simple linearisation causes considerably inaccurate approximation. To address these limitations, alternative approaches, such as Gaussian higher-order Kalman filters (GHKFs) that take second- or even higher-order approximation of the nonlinearity, are proposed~\cite{gelb1974applied}. Another set of alternatives are sigma point Kalman filters (SPKF), including the unscented Kalman filter (UKF)~\cite{julier2004unscented}, the central difference filter~\cite{ito2000gaussian}, the divided difference filter~\cite{norgaard2000new}, the cubature Kalman filter~\cite{arasaratnam2009cubature}, the conjugate unscented Kalman filter~\cite{adurthi2017conjugate}, etc. SPKFs essentially are sampling-based filters and can yield more accurate results than EKFs or GHKFs which are both derivative dependent, especially when accurate initial condition states are poorly known~\cite{crassidis2006sigma}. The Gaussian sum Kalman filter (GSKF) is also proposed to approximate a posterior non-Gaussian probability density function (PDF) of the state using a weighted sum of Gaussian PDFs \cite{ito2000gaussian,yang2017analysis}. For the general OD problem, if measurements are subject to gross errors or outliers, which is the case of optical tracking usually, non-Gaussian behaviours due to both the orbital state and measurements have to be handled in the sequential filtering framework. The characteristics of these non-Gaussian PDFs can be accurately captured by introducing higher-order moments such as skewness and kurtosis, as seen in the high-order unscented estimator (HOUSE)~\cite{stojanovski2021house}. 
Beyond addressing non-Gaussinity, the nonlinearity inherent in orbit propagation can also be mitigated by using orbital elements, such as the modified equinoctial orbital elements (MEEs)~\cite{walker1985modified}, instead of Cartesian coordinates. This format has demonstrated the capability to reduce the nonlinearity of orbital dynamics over more extended time periods~\cite{psiaki2022gaussian,yun2022kernel}, which, in turn, curtails the growth of non-Gaussianity due to the nonlinearity of the orbit propagator.

This work aims to tackle the nonlinear and non-Gaussian OD problem by introducing a novel version of HOUSE, dubbed $w$-HOUSE. $w$-HOUSE leverages a square-root formulation for covariance evolution, eliminating the requirement of high kurtosis values for the state to be estimated. To assess the efficacy of $w$-HOUSE, comparisons are made with other filtering methods, including UKF, square-root UKF (SRUKF), two conjugate unscented transformation (CUT) filters, i.e., CUT-4 and CUT-6, and $\delta$-HOUSE. 
The evaluation encompasses three numerical examples: a projectile example inherited from the previous work~\cite{stojanovski2021house}, a ground-based radar tracking scenario, and a real-world ground-based optical tracking scenario.

This paper makes several significant contributions to the OD field: 
\begin{itemize}
    \item \textcolor{red}{It introduces a new algorithm, $w$-HOUSE, which is based on a square-root formulation. This new approach eliminates the need for high kurtosis in the state being estimated, a requirement of the original $\delta$-HOUSE. As a result, $w$-HOUSE exhibits increased robustness with respect to the varying control parameter. In other words, $w$-HOUSE is more resilient to changes in the $w$ parameter compared to $\delta$-HOUSE's sensitivity to changes in the $\delta$ parameter.}
    \item It presents a ground-based angle-only OD algorithm that achieves state-of-the-art performance in terms of accuracy and efficiency, even with outlier-contaminated measurements. An orbit propagator based on MEEs is employed across all filters. The proposed $w$-HOUSE outperforms the UKF, CUT-4, and CUT-6 filters in terms of positioning errors, achieving better than \SI{60}{\meter} of three-dimensional (3D) root mean square errors (RMSE) in a three-day scenario. Its runtime is comparable to UKFs and significantly more efficient than CUT filters.
    \item It provides an OD toolkit written in C++ and will be released to public along the publication of this paper. This toolkit offers the community valuable insights into advanced filtering techniques for the OD problem, a fundamental challenge for numerous satellite missions and space situational awareness applications.
\end{itemize}
The rest of the paper is organised as follows. Section~\ref{sec:bayesfiltering} introduces the fundamental Bayesian theorem as applied to the specific OD problem, where orbit dynamics and measurement models are presented. Section~\ref{sec:filters} revisits the sigma point sampling based linear least mean squares (LLMS) filters, including UKF, CUT filters, and HOUSE, and particular emphasis is placed on the original $\delta$-HOUSE. The novel $w$-HOUSE method is introduced by leveraging the square-root formulation in Section~\ref{sec:srhouse}. Section~\ref{sec:simulations} presents numerical examples, including a projectile example inherited from prior work, as well as two OD scenarios featuring synthetic and real-world measurements. Results and analyses of the OD outcomes are provided in Section~\ref{sec:results}, with concluding remarks to follow in Section~\ref{sec:conclusion}.

\section{Bayesian Filtering for Orbit Determination}
\label{sec:bayesfiltering}

\subsection{Systematic Equations with Initial Conditions}
Consider the discrete-time state space models~\cite{sachin2012nonlinear}:
\begin{gather}
    \vec{\boldsymbol{x}}(k) = \boldsymbol{F}(\vec{\boldsymbol{x}}(k-1),\vec{\boldsymbol{\omega}}(k-1)),
    \label{eq:dyn_system}
\end{gather}
\begin{gather}    
    \vec{\boldsymbol{z}}(k) = \boldsymbol{h}(\vec{\boldsymbol{x}}(k),\vec{\boldsymbol{\nu}}(k)),
    \label{eq:meas_system}
\end{gather}
where $\vec{\boldsymbol{x}}(k-1)$ indicates the orbital state at time $t_{k-1}$; $\vec{\boldsymbol{z}}(k)$ is the measurement at time $t_{k}$; $\vec{\boldsymbol{\omega}}$ and $\vec{\boldsymbol{\nu}}$ represent the noise associated with the orbital state and measurement variables, respectively, both of which follow a specific PDF such as a Gaussian PDF and a non-Gaussian Pearson Type PDF~\cite{stojanovski2021house}; $\boldsymbol{F}$ represents the nonlinear dynamics driving the discrete-time evolution of the states; and $\boldsymbol{h}$ is the nonlinear measurement function. The dimension of the state and measurement are represented by $n_x$ and $n_z$, respectively. It is assumed that $\vec{\boldsymbol{\omega}}$ and $\vec{\boldsymbol{\nu}}$ are independent of $\vec{\boldsymbol{x}}$. 

Given the known PDFs $p(\vec{\boldsymbol{\omega}})$, $p(\vec{\boldsymbol{\nu}})$, $p_{0|0}(\vec{\boldsymbol{x}})$ for the stochastic processes $\vec{\boldsymbol{\omega}}(k)$, $\vec{\boldsymbol{\nu}}(k)$ and the initial state $\vec{\boldsymbol{x}}(0)$, respectively, the objective of the sequential Bayesian filtering for OD is to find the PDF $p_{k|k}(\vec{\boldsymbol{x}})$ from $p_{k|k-1}(\vec{\boldsymbol{x}})$ given $p_{k}(\vec{\boldsymbol{z}}|\vec{\boldsymbol{x}})$ in the following Eq.~\ref{eq:eq_meas_update}, so that the estimate of the orbital state can be obtained. The operation of the filter can be divided into two recursive steps: time update and measurement update. During the time update step, the objective is to determine the PDF of $\vec{\boldsymbol{x}}(k|k-1)$ given the PDFs of $\vec{\boldsymbol{x}}(k-1)$ and $\vec{\boldsymbol{\omega}}(k-1)$. During the measurement update step, the objective is to determine the PDF of $\vec{\boldsymbol{x}}(k)$ given a measurement $\vec{\boldsymbol{z}}(k)$, the PDF of $\vec{\boldsymbol{\nu}}(k)$, and the prior predicted PDF of $\vec{\boldsymbol{x}}(k|k-1)$. This step refines the state PDF by incorporating the measurement information and adjusting the predicted PDF accordingly. By alternating between these two steps, the filter recursively updates and refines the state estimate as new measurements become available, improving the accuracy of the estimated state over time.

\subsection{Time Update for $t\in[t_{k-1},t_{k}],\ k\texorpdfstring{\geq}{>=} 1$}
Following the most general form by the Chapman-Kolmogorov equation, the time update step involves propagating the state PDF forward in time, so the prior PDF at $t_k$ is given by: 
\begin{gather}
    p_{k|k-1}(\vec{\boldsymbol{x}}) = \int p_{k|k-1}(\vec{\boldsymbol{x}}|\vec{\boldsymbol{x}^\prime})p_{k-1}(\vec{\boldsymbol{x}^\prime})\text{d} \vec{\boldsymbol{x}^\prime},
\end{gather}
where $p_{k|k-1}(\vec{\boldsymbol{x}})$ is the predicted or prior PDF for the orbial state $\vec{\boldsymbol{x}}(k)$, and $p_{k|k-1}(\vec{\boldsymbol{x}}|\vec{\boldsymbol{x}^\prime})$ is the conditional PDF for $\vec{\boldsymbol{x}}(k)$ given $\vec{\boldsymbol{x}}(k-1)=\vec{\boldsymbol{x}^\prime}$. The PDF $p_{k-1}(\vec{\boldsymbol{x}^\prime})$ is the prior PDF for $\vec{\boldsymbol{x}}(k-1)$, which is obtained from the previous measurement update step or the initial value for the first time epoch. The PDF $p_{k}(\vec{\boldsymbol{x}})$ is then used in the measurement update step. According to Eq.~\ref{eq:dyn_system}, the transition from $\vec{\boldsymbol{x}}(k-1)$ to $\vec{\boldsymbol{x}}(k)$ is solely affected by the process noise $\vec{\boldsymbol{\omega}}(k-1)$. Hence, the PDF $p_{k|k-1}(\vec{\boldsymbol{x}}|\vec{\boldsymbol{x}^\prime})$ can be written based on a Dirac delta function ($\delta_{\text{Dirac}}$):
\begin{gather}
    p_{k|k-1}(\vec{\boldsymbol{x}}|\vec{\boldsymbol{x}^\prime}) = \int \delta_{\text{Dirac}}(\vec{\boldsymbol{x}}-\boldsymbol{F}(\vec{\boldsymbol{x}^\prime}, \vec{\boldsymbol{\omega}}))p_{k-1}(\vec{\boldsymbol{\omega}})\text{d} \vec{\boldsymbol{\omega}}.
\end{gather}

There are various ways of arriving at the discrete form of the orbit dynamics equation above. For example, Eq.~\ref{eq:dyn_system} can be derived from a continuous-time model of orbit dynamics in the Cartesian coordinates:
\begin{gather}
    \frac{\text{d}}{\text{d}t}\vec{\boldsymbol{x}}(t) = \boldsymbol{f}(\vec{\boldsymbol{x}}(t),\vec{\boldsymbol{\omega}}(t)),
    \label{eq:dyn_system_con}
\end{gather}
where ${\vec{\boldsymbol{x}}}$ comprises the position ${\vec{\boldsymbol{r}}}$ and ${\vec{\boldsymbol{v}}}$ vectors. Usually these orbital states are represented in the Earth centred inertial (ECI) system. For satellites in the near earth orbit, gravitational and non-gravitational perturbations have to be accounted for. In general the accelerations acting on satellites consist of terms for the Earth gravitational force, the 3rd-body gravitational attraction of the Sun and Moon, the solar radiation pressure and atmosphere drag on the satellite, if no active orbital manoeuvre is performed. The exact formulations for each term included in $\boldsymbol{f}(\vec{\boldsymbol{x}})$ can be obtained from classical books, for example,~\cite{montenbruck2012satellite} or the author's previous work~\cite{yang2022phifa}. 
Given a sampling interval $T$, it is assumed the variation of the unmeasured disturbances can be adequately approximated using piecewise constant functions of the form $\vec{\boldsymbol{\omega}}(t) = \vec{\boldsymbol{\omega}}(k),\ \text{for}\ t\in[t_{k-1},t_{k}]$. As a result, the discrete-time model in Eq.~\ref{eq:dyn_system} can be obtained by integrating Eq.~\ref{eq:dyn_system_con} over the interval $[t_{k-1},t_{k}]$ ($t_{k}=t_{k-1}+T$):
\begin{gather}
    \label{eq:discrete_time_model}
    \boldsymbol{F}(\vec{\boldsymbol{x}}(k-1),\vec{\boldsymbol{\omega}}(k-1)) =  \nonumber \\\vec{\boldsymbol{x}}(k-1) + \int_{(k-1)T}^{kT} \boldsymbol{f}(\vec{\boldsymbol{x}}(\tau),\vec{\boldsymbol{\omega}}(k))\text{d}\tau.
\end{gather} 

Instead of formulating the orbital dynamics in the Cartesian coordinate system, orbital elements are used to describe the orbital dynamics and present advantageous features in eliminating non-linearity. It is favourable for handling orbital uncertainty propagation within a filtering framework. The MEE is introduced additionally to represent the motion of RSOs with respect to the ECI frame~\cite{walker1985modified,das2019optimal}. The state vector is $\vec{\boldsymbol{x}}_{\text{mee}} = (p,f,g,h,k,L)^T$, where the first five elements are real numbers while the last one is a circular variable. The MEE is defined via the classical orbital elements (COE) $\vec{\boldsymbol{x}}_{\text{coe}} = (a,e,i,\omega,\Omega,\nu)^T$:
\begin{equation}
\begin{aligned}
    p &= a (1 - e^2),\\
	f &= e \cos(\omega + \Omega),\\
	g &= e \sin(\omega + \Omega),\\
	h &= \tan(i / 2) \cos(\Omega),\\
	k &= \tan(i / 2) \sin(\Omega),\\
	L &= \Omega + \omega + \nu,
    \label{eq:mee_from_coe}
\end{aligned}
\end{equation}
where $p$ is semiparameter, $a$ is semi-major axis, $e$ is orbital eccentricity, $i$ is the orbital inclination, $\omega$ is argument of perigee, $\Omega$ is right ascension of the ascending node, $\nu$ is true anomaly, and $L$ is true longitude. There are no singularities with zero eccentricity and inclination by using MEE, except when an orbit has an inclination of $\pm 180^\circ$. These extreme retrograde orbits are rare and can be ignored in this work. As a result, the following ordinary differential equations are used to describe the orbital dynamics,
\begin{figure*}
    \begin{equation}
    \begin{aligned}
        \label{eq:mee_eom}
        \dot{p} &= \frac{2p}{q}\sqrt{\frac{p}{\mu}}a_t, \\
        \dot{f} &= \sqrt{\frac{p}{\mu}} \Bigg\{\frac{[(1+q)\cos(L) + f]a_t - g(h\sin(L) - k\cos(L))a_n}{q} + a_r\sin(L) \Bigg\},\\
        \dot{g} &= \sqrt{\frac{p}{\mu}} \Bigg\{\frac{[(1+q)\sin(L) + g]a_t + f(h\sin(L) - k\cos(L))a_n}{q}-a_r\cos(L) \Bigg\},\\
        \dot{h} &= \sqrt{\frac{p}{\mu}}\frac{s^2a_n}{2q}\cos(L),\\
        \dot{k} &= \sqrt{\frac{p}{\mu}}\frac{s^2a_n}{2q}\sin(L),\\
        \dot{L} &= \sqrt{\mu p}\frac{z^2}{p^2} + \sqrt{\frac{p}{\mu}}\frac{(h\sin(L)-k\cos(L))a_n}{q},
    \end{aligned}
    \end{equation}
\end{figure*}
where $s^2=1+h^2+k^2$, $q = 1+f\cos(L)+g\sin(L)$, and $a_r$, $a_t$ and $a_n$ are non-two-body acceleration components due to perturbations in the radial, tangential and normal directions, respectively. The radial direction corresponds to the geocentric radius vector of the RSO, measured positively in an outward direction from the Earth centre. The tangential direction is perpendicular to this radius vector and measured positively in the direction of orbital motion. The normal direction is positive along the angular momentum vector of the RSO's orbit.

\subsection[Measurement Update]{Measurement Update for $t=t_{k}$}
The prior PDF is then updated via Bayesian theorem once a measurement is available:
\begin{gather}
    p_{k|k}(\vec{\boldsymbol{x}}) = \frac{p_{k}(\vec{\boldsymbol{z}}|\vec{\boldsymbol{x}})p_{k|k-1}(\vec{\boldsymbol{x}})}{\int p_{k}(\vec{\boldsymbol{z}}|\vec{\boldsymbol{x}})p_{k|k-1}(\vec{\boldsymbol{x}})}, 
    \label{eq:eq_meas_update}
\end{gather}
where $p_{k}(\vec{\boldsymbol{z}}|\vec{\boldsymbol{x}})$ is the measurement likelihood function given by 
\begin{gather}
    p_{k}(\vec{\boldsymbol{z}}|\vec{\boldsymbol{x}})=\int \delta_{\text{Dirac}}(\vec{\boldsymbol{z}}-\boldsymbol{h}(\vec{\boldsymbol{x}}(k),\vec{\boldsymbol{\nu}}(k)))p_{\nu}\text{d}\vec{\boldsymbol{\nu}}.
    \label{eq:eq_meas_lh}
\end{gather}
Normally there is no analytical solution to the integral in Eq.~\,\ref{eq:eq_meas_update}. Moments are calculated to approximate the prior and posterior PDFs of the state, which results in various filters.

Based on a traditional astronomical telescope, measurements of topocentric right ascension $\alpha_{\text{RA}}$ and declination $\delta_{\text{Dec}}$ can be obtained. They are defined in the topocentric coordinate system and are calculated from Cartesian coordinates of the RSO and tracking sensor as follows: 
\begin{gather} \label{eq_angularmeasmodel}
    \alpha_{\text{RA}} = \tan^{-1} \Big( \frac{y_{\text{rso}}-y_{\text{stn}}}{x_{\text{rso}}-x_{\text{stn}}} \Big), \\
    \delta_{\text{Dec}}  = \sin^{-1} \Big( \frac{z_{\text{rso}}-z_{\text{stn}}}{\rho} \Big),
\end{gather}
where $\rho$ is the range between the RSO (indicated by the subscript "$\text{rso}$") and the sensor (indicated by the subscript "$\text{stn}$"). A Radar sensor can measure the range and the range rate as additional two observables shown as below: 
\begin{gather} \label{eq_radarmeasmodel}
    \rho = \sqrt{(x_{\text{rso}}-x_{\text{stn}})^2 + (y_{\text{rso}}-y_{\text{stn}})^2 + (z_{\text{rso}}-z_{\text{stn}})^2},  \\
    \dot \rho  =\frac{\vec{\boldsymbol{r}}_{\text{rso}} \cdot \vec{\boldsymbol{v}}_{\text{rso}}}{\rho},
\end{gather}
where $\vec{\boldsymbol{r}}_{\text{rso}}$ and $\vec{\boldsymbol{v}}_{\text{rso}}$ are vectors of position and velocity, respectively. When the orbit is propagated via MEE, a transformation is necessary between MEE to the Cartesian coordinate. Hence, the aforementioned measurement equations can be rewritten w.r.t. MEE easily so are skipped here. 

\section{Sigma Point Sampling Based Linear Least Mean Squares Filters}
\label{sec:filters}
To obtain analytical solutions to the sequential Bayesian filtering for OD introduced in Section~\ref{sec:bayesfiltering}, one type of methods is to approximate the posterior PDF given by Eq.~\ref{eq:eq_meas_update} and apply one optimisation criteria to obtain the optimal estimates from the posterior PDF~\cite{sachin2012nonlinear}. Minimising the mean square error is a widely used criteria which was first developed by Kalman~\cite{kalman1960new} for linear systems. This has been extended to nonlinear systems by using various ways, for example, approximation based on Taylor series expansion and LLMS approximation.  
LLMS approximation is a technique to approximate the distribution of a nonlinear function of a random variable by linearising it around its expected value~\cite{gelb1974applied,soderstrom2002discrete}, which is shown as a better alternative for approximating the nonlinear function than employing the Taylor expansion based linearisation. LLMS estimators can work with Eqns.~\ref{eq:dyn_system}-\ref{eq:meas_system} directly and can also be applied for non-Gaussian PDFs if these PDFs are known. 

\subsection{Sigma Point Sampling Based LLMS Estimators}
Sigma point sampling based LLMS estimators, also known as sigma point filters, are a class of LLMS estimators that use sigma points to approximate moments of a PDF, such as mean, variance, skewness, and kurtosis. These moments are essential for characterising the shape and properties of the PDF. Different choices of sigma points lead to different sigma point filters as introduced in Section~\ref{sec:introduction}. However, the general concept in these filters is sets of predicted state values and measurement values are used to calculate the estimated covariance matrices, which are then used to calculate the Kalman gain matrix in a linear manner. Hence, the optimal state estimate is constructed recursively by combining the predicted estimate with measurements. These steps are summarised in the Appendix. 

The UKF, proposed by Julier and Uhlmann~\cite{julier1997new}, uses an unscented transformation (UT) to approximate the state PDF. It is based on the assumption that the state follows a Gaussian distribution. The nonlinear UT employs sigma points generated by using Eq.~\ref{eq:sig_ut} to approximate the first two orders of moments, i.e., mean and covariance, of the state:
\begin{gather}
    {\vec{\boldsymbol{x}}}^j = \mu + \sqrt{n+\kappa}\boldsymbol{c}^j\quad w^j = \frac{1}{2(n+\kappa)},\notag\\
    {\vec{\boldsymbol{x}}}^{j+n} = \mu - \sqrt{n+\kappa}\boldsymbol{c}^j\quad w^{j+n} = \frac{1}{2(n+\kappa)},\label{eq:sig_ut}\\
    {\vec{\boldsymbol{x}}}^{2n+1} = \mu\quad w^{2n+1} = \kappa/(n+\kappa),\notag
\end{gather}
where $\mu$ is the mean of the $n$-dimensional state variable ${\vec{\boldsymbol{x}}}$, the $\boldsymbol{c}^j$ is the $j^{\text{th}}$ column of original Cholesky factor of ${\boldsymbol{P}_{x}}$, $\kappa$ is a tuning parameter, and $w$ is the corresponding weight. A tutorial of the original UKF algorithm is summarised in~\cite{terejanu2011unscented}, and  its square-root form is presented in~\cite{merwe2001square}.

Besides first and second orders of moments, Adurthi et al.~\cite{adurthi2017conjugate} proposed a unified approach, labelled as CUT, to generate sigma points that approximate higher-order moments for Gaussian and uniform PDFs. These filters are called CUT filters. CUT introduces additional axes, labeled as conjugate axes, such that the resultant points are more "spread out" symmetrically in space. In contrast, the UT selects sigma points that lie on carefully selected axes, such as the principal axes or the eigenvectors of the covariance matrix. The weights for the CUT and UT are calculated using different formulas, which are derived based on the moment matching criteria. Therefore, the CUT and UT differ in the selection of sigma points and corresponding weights, which can affect the accuracy and efficiency of the approximation.

Still considering high-order moments, such as the state skewness and kurtosis besides the mean and covariance, Stojanovski and Savransky~\cite{stojanovski2021house} extended the UKF based on the method by Ponomareva et al.~\cite{ponomareva2010new}. The generation of sigma points is based on a more general form shown in Eq.~\ref{eq:sig_house}:
\begin{gather}
    {\vec{\boldsymbol{x}}}^j = \mu + \vec{\boldsymbol{\alpha}}\circ\boldsymbol{c}^j\quad w^j = \frac{1}{\vec{\boldsymbol{\alpha}}\cdot(\vec{\boldsymbol{\alpha}} + \vec{\boldsymbol{\beta}})n},\notag\\
    {\vec{\boldsymbol{x}}}^{j+n} = \mu - \vec{\boldsymbol{\beta}}\circ\boldsymbol{c}^j\quad w^{j+n} = \frac{1}{\vec{\boldsymbol{\beta}}\cdot(\vec{\boldsymbol{\alpha}} + \vec{\boldsymbol{\beta}})n},\label{eq:sig_house}\\
    {\vec{\boldsymbol{x}}}^{2n+1} = \mu\quad w^{2n+1} = 1 - \sum_{j=1}^{2n}w^j.\notag
\end{gather}
where $\vec{\boldsymbol{\alpha}}$ and $\vec{\boldsymbol{\beta}}$ are coefficient vectors chosen to preserve the marginal third-and fourth-order moments of $\vec{\boldsymbol{x}}$ and can be calculated via the following Eqns.~\ref{eq:weight_alpha} and \ref{eq:weight_beta}, and the operator $\circ$ indicates element-wise multiplication between two vectors. The resulted filter is called the Higher-Order Unscented Estimator~\cite{stojanovski2021house}. It is designed to be more robust in cases where the initial distribution, process noise and measurement noise have high kurtosis. It is also indicated in~\cite{stojanovski2021house} that HOUSE is computationally more efficient than the fourth and higher-order CUT filters for three given simulation examples. 

\subsection{High-Order Unscented Estimator}
HOUSE accounts for non-Gaussian and nonuniform distributions by propagating the state skewness and kurtosis in addition to the mean and covariance. Skewness represents the asymmetry of the PDF, indicating whether the probability mass is concentrated more to the left or right of the mean. Kurtosis, on the other hand, measures the heaviness of the tails of the distribution, highlighting the presence of extreme values, e.g., outliers associated with measurements. Explicit solutions for sample points and weights can be obtained via Eq.~\ref{eq:sig_house}, making evaluations via HOUSE easier than CUT filters. However, the original HOUSE filter suffers from possible negative value of the last weight given by Eq.~\ref{eq:sig_house}. The solution proposed in~\cite{stojanovski2021house} is to introduce a $\delta$ parameter to intentionally increase the small values of kurtosis. The working procedures of the original $\delta$-HOUSE are introduced as below. 

\subsubsection{Time Update}
The time update process begins with the generation of sigma points. This is achieved by considering the distribution of the augmented state vector, which is represented by its mean:
\begin{gather}
    {\vec{\boldsymbol{x}}_{a}}(k-1) = \begin{bmatrix}
        {\vec{\boldsymbol{x}}}_m(k-1)\\
        {\vec{\boldsymbol{0}}_{\omega}}(k-1)
    \end{bmatrix},
    \label{eq:aug_pred_state}
\end{gather}
its covariance:
\begin{gather}
    {\boldsymbol{P}}_a(k-1) = \begin{bmatrix}
        {\boldsymbol{P}}_x(k-1) & \boldsymbol{0}\\
        \boldsymbol{0} & {\boldsymbol{P}}_{\omega}(k-1)
    \end{bmatrix},
    \label{eq:aug_pred_covariance}
\end{gather}
its skewness:
\begin{gather}
    {\vec{\boldsymbol{\gamma}}_a}(k-1) = \begin{bmatrix}
        {\vec{\boldsymbol{\gamma}}}_x(k-1)\\
        {\vec{\boldsymbol{\gamma}}}_{\omega}(k-1)
    \end{bmatrix},
    \label{eq:aug_pred_skewness}
\end{gather}
and its kurtosis:
\begin{gather}
    {\vec{\boldsymbol{\kappa}}_a}(k-1) = \begin{bmatrix}
        {\vec{\boldsymbol{\kappa}}}_x(k-1)\\
        {\vec{\boldsymbol{\kappa}}}_{\omega}(k-1)
    \end{bmatrix},
\end{gather}
where $k$ indicates the $k^{\text{th}}$ time epoch and $k\geq 1$, and $\vec{\boldsymbol{0}}_{\omega}$ is an $n_{\omega}$-dimensional zero vector. This results in $2(n_x+n_{\omega})+1$ sigma points, which are represented as:
\begin{gather}
    {\vec{\boldsymbol{x}}}^{(j)}_a(k-1) = \begin{bmatrix}
        {\vec{\boldsymbol{x}}}^{(j)}(k-1)\\
        {\vec{\boldsymbol{\omega}}}^{(j)}(k-1)
    \end{bmatrix},
\end{gather}
where $1\leq j\leq 2(n_x+n_{\omega})+1$. The calculation of the vector of weights ${\vec{\boldsymbol{w}}}$\footnote{Note the weight vector $\vec{\boldsymbol{w}}$ is different from the process noise vector ${\vec{\boldsymbol{\omega}}}(k)$.} for all sigma points can be referred to the Section III.B of~\cite{stojanovski2021house} and is also presented in the following Eqns.~\ref{eq:sig_srhouse}-\ref{eq:weight_beta}. To guarantee the nonnegativeness of the last weight $w^{2n+1}$, one way is to increase the small values of kurtosis. For any component $\kappa_i$, if it is smaller than $\frac{n}{1-\delta}+\gamma_i^2$, let, 
\begin{gather}
    \kappa_i = \frac{n}{1-\delta}+\gamma_i^2,
    \label{eq:upd_kappa}
\end{gather}
where $\delta$ is a predefined \textcolor{red}{control parameter}, $n$ is the dimension of the random variable, i.e., the state ${\vec{\boldsymbol{x}}}$ or the process noise ${\vec{\boldsymbol{\omega}}}$, and $\gamma_i$ is the corresponding skewness component. \textcolor{red}{As a result, the last weight 
\begin{gather}
    w^{2n+1} = \delta,
    \label{eq:weight_last}
\end{gather}
and its nonnegativeness is guaranteed by defining a nonnegative parameter of $\delta$}.

The next phase involves the propagation of each sigma point using the orbital dynamics models:
\begin{gather}
    {\vec{\boldsymbol{x}}}_a^{(j)}(k|k-1) = \boldsymbol{F}\big(\vec{\boldsymbol{x}}^{(j)}_a(k-1)\big).
\end{gather}

Subsequently, the predicted mean and covariance are computed as follows: 
\begin{gather}
    {\vec{\boldsymbol{x}}}_{a,m}(k|k-1)=\sum_{j=1}^{2(n_x+n_{\omega})+1} w^{(j)}\vec{\boldsymbol{x}}_{a}^{(j)}(k|k-1),   
\end{gather}
\begin{gather}
    \vec{\delta\boldsymbol{x}}_{a}^{(j)}(k|k-1) =\vec{\boldsymbol{x}}_{a}^{(j)}(k|k-1) - {\vec{\boldsymbol{x}}}_{a,m}(k|k-1),
    \label{eq:pred_deviations}
\end{gather}
\begin{gather}
    \begin{split}
        {\boldsymbol{P}}_{x}(k|k-1) = \sum_{j=1}^{2(n_x+n_{\omega})+1} w^{(j)}\vec{\delta\boldsymbol{x}}^{(j)}(k|k-1) \\
        \times \Big[\vec{\delta\boldsymbol{x}}^{(j)}(k|k-1)\Big]^T,
    \end{split}
    \label{eq:pred_covariance}
\end{gather}
where $\vec{\delta\boldsymbol{x}}^{(j)}(k|k-1)$ is the first $n_x$ rows of $\vec{\delta\boldsymbol{x}}_{a}^{(j)}(k|k-1)$.

Following this, the normalised states for the sigma points are computed:
\begin{gather}
    \vec{\tilde{\boldsymbol{x}}}^{(j)}(k|k-1) = \Big(\sqrt{{\boldsymbol{P}}_x(k|k-1)}\Big)^{-1} \vec{\delta\boldsymbol{x}}^{(j)}(k|k-1). 
    \label{eq:pred_normalised_state}  
\end{gather}

Finally, the skewness and kurtosis of the normalised state are computed:
\begin{gather}
    \vec{\boldsymbol{\gamma}}_{x}(k|k-1)  = \sum_{j=1}^{2(n_{x} + n_{\omega})+1} w^{(j)}\Big[\vec{\tilde{\boldsymbol{x}}}^{(j)}(k|k-1)\Big]^3,    
    \label{eq:pred_skewness}
\end{gather}
\begin{gather}
    \vec{\boldsymbol{\kappa}}_{x}(k|k-1) = \sum_{j=1}^{2(n_{x} + n_{\omega})+1} w^{(j)}\Big[\vec{\tilde{\boldsymbol{x}}}^{(j)}(k|k-1)\Big]^4.    
    \label{eq:pred_kurtosis}
\end{gather}

\subsubsection{Measurement Update}
The measurement update process begins by generating sigma points. This is achieved by considering the distribution of the augmented state vector, which is represented by its mean:
\begin{gather}
    {\vec{\boldsymbol{x}}}_a(k) = \begin{bmatrix}
        {\vec{\boldsymbol{x}}_m}(k|k-1)\\
        {\vec{\boldsymbol{0}}_{\nu}}(k)
    \end{bmatrix},
    \label{eq:aug_corr_state}
\end{gather}
its covariance:
\begin{gather}
    {{\boldsymbol{P}}_{a}}(k) = \begin{bmatrix}
        {\boldsymbol{P}}_x(k|k-1) & \boldsymbol{0}\\
        \boldsymbol{0} & {\boldsymbol{P}}_{\nu}(k)
    \end{bmatrix},
    \label{eq:aug_corr_covariance}
\end{gather}
its skewness:
\begin{gather}
    {\vec{\boldsymbol{\gamma}}_{a}}(k) = \begin{bmatrix}
        {\vec{\boldsymbol{\gamma}}}_x(k|k-1)\\
        {\vec{\boldsymbol{\gamma}}}_{\nu}(k)
    \end{bmatrix},
    \label{eq:aug_corr_skewness}
\end{gather}
and its kurtosis:
\begin{gather}
    {\vec{\boldsymbol{\kappa}}_{a}}(k) = \begin{bmatrix}
        {\vec{\boldsymbol{\kappa}}}_x(k|k-1)\\
        {\vec{\boldsymbol{\kappa}}}_{\nu}(k)
    \end{bmatrix},
    \label{eq:aug_corr_kurtosis}
\end{gather}
where ${\vec{\boldsymbol{x}}_m}(k|k-1)$, ${\boldsymbol{P}_x}(k|k-1)$, ${\vec{\boldsymbol{\gamma}}_x}(k|k-1)$ and ${\vec{\boldsymbol{\kappa}}_x}(k|k-1)$ can be extracted from the corresponding augmented vectors and matrix in the time update, and $\vec{\boldsymbol{0}}_{\nu}$ is an $n_{\nu}$-dimensional zero vector. This results in $2(n_x+n_{\nu})+1$ sigma points, which are represented as:
\begin{gather}
    {\vec{\boldsymbol{x}}}^{(j)}_{a}(k) = \begin{bmatrix}
        {\vec{\boldsymbol{x}}}^{(j)}(k)\\
        {\vec{\boldsymbol{\nu}}}^{(j)}(k)
    \end{bmatrix},
\end{gather}
where $1\leq j\leq 2(n_x+n_{\nu})+1$. The next phase involves the calculation of the corresponding measurement for each sigma point using the measurement model:
\begin{gather}
    {\vec{\boldsymbol{z}}}^{(j)}(k) = \boldsymbol{h}\big(\vec{\boldsymbol{x}}^{(j)}_{a}(k)\big).
\end{gather}
Subsequently, the mean of sigma points of the augmented state and the deviations of sigma points from the mean are calculated as:
\begin{gather}
    {\vec{\boldsymbol{x}}}_{a,m}(k)=\sum_{j=1}^{2(n_x+n_{\nu})+1} w^{(j)}\vec{\boldsymbol{x}}_{a}^{(j)}(k),     
\end{gather}
\begin{gather}
    \vec{\delta\boldsymbol{x}}_{a}^{(j)}(k) =\vec{\boldsymbol{x}}_a^{(j)}(k) - {\vec{\boldsymbol{x}}}_{a,m}(k).
    \label{eq:corr_deviations}    
\end{gather}
Following this, the measurement mean and covariance are computed:
\begin{gather}
    {\vec{\boldsymbol{z}}}_m(k)=\sum_{j=1}^{2(n_{x} + n_{\nu})+1} w^{(j)}\vec{\boldsymbol{z}}^{(j)}(k),     
\end{gather}
\begin{gather}
    \vec{\delta\boldsymbol{z}}^{(j)}(k) =\vec{\boldsymbol{z}}^{j}(k) - {\vec{\boldsymbol{z}}}_m(k),
    \label{eq:corr_meas_deviations}
\end{gather}
\begin{gather}
    {\boldsymbol{P}}_{z}(k)=\sum_{j=1}^{2(n_{x} + n_{\nu})+1} w^{(j)}\vec{\delta\boldsymbol{z}}^{(j)}(k)\Big[\vec{\delta\boldsymbol{z}}^{(j)}(k)\Big]^T, 
    \label{eq:corr_meas_covariance}    
\end{gather}
\begin{gather}
    {\boldsymbol{P}}_{xz}(k)=\sum_{j=1}^{2(n_{x} + n_{\nu})+1} w^{(j)}\vec{\delta\boldsymbol{x}}^{(j)}(k)\Big[\vec{\delta\boldsymbol{z}}^{(j)}(k)\Big]^T,
    \label{eq:corr_cross_covariance}  
\end{gather}
where $\vec{\delta\boldsymbol{x}}^{(j)}(k)$ is the first $n_x$ rows of $\vec{\delta\boldsymbol{x}}_{a}^{(j)}(k)$. 
The Kalman gain is then calculated as:
\begin{gather}
    {\boldsymbol{K}}(k)={\boldsymbol{P}}_{xz}(k){\boldsymbol{P}}_{z}^{-1}(k).
    \label{eq:corr_kalman_gain}
\end{gather}
The mean and covariance are updated using the following linear forms:
\begin{gather}
    {\vec{\boldsymbol{x}}}_m(k)={\vec{\boldsymbol{x}}}_m(k|k-1)+{\boldsymbol{K}}(k)(\vec{\boldsymbol{z}}(k) - {\vec{\boldsymbol{z}}}_m(k)),
    \label{eq:corr_mean}
\end{gather}
\begin{gather}
    {\boldsymbol{P}}_{x}(k)={\boldsymbol{P}}_{x}(k|k-1)-{\boldsymbol{K}}(k){\boldsymbol{P}}_{xz}^T(k).
    \label{eq:corr_covariance}
\end{gather}
Based on the deviations in Eq.~\ref{eq:corr_deviations}, normalised sigma points are calculated:
\begin{gather}
    \vec{\tilde{\boldsymbol{x}}}^{(j)}(k) = \Big(\sqrt{{\boldsymbol{P}}_{x}(k)}\Big)^{-1} \vec{\delta\boldsymbol{x}}^{(j)}(k).
    \label{eq:corr_normalised_state}  
\end{gather}
Accordingly, the skewness and kurtosis vectors are computed as:
\begin{gather}
    \vec{\boldsymbol{\gamma}}_{x}(k) = \sum_{j=1}^{2(n_{x} + n_{\nu})+1} w^{(j)}\Big[\vec{\tilde{\boldsymbol{x}}}^{(j)}(k)\Big]^3,    
    \label{eq:corr_skewness}
\end{gather}
and 
\begin{gather}
    \vec{\boldsymbol{\kappa}}_{x}(k) = \sum_{j=1}^{2(n_{x} + n_{\nu})+1} w^{(j)}\Big[\vec{\tilde{\boldsymbol{x}}}^{(j)}(k)\Big]^4.
    \label{eq:corr_kurtosis} 
\end{gather}
The time update and measurement update are running recursively from the first time epoch till the last one. The $\delta$-HOUSE algorithm is summarised in Algorithm~\ref{algo:HOUSE}.

\begin{algorithm*}
    \justifying
    \caption{$\delta$-HOUSE Algorithm}
    \label{algo:HOUSE}
    \KwData{Initial time epoch, initial state PDF represented by the mean $\vec{\boldsymbol{x}}_m(0)$, covariance $\textbf{cholesky}({\boldsymbol{P}}_{x}(0))$, skewness $\vec{\boldsymbol{\gamma}}_{x}(0)$ and kurtosis $\vec{\boldsymbol{\kappa}}_{x}(0)$, the process noise PDF represented by the covariance ${\boldsymbol{Q}}$, skewness $\vec{\boldsymbol{\gamma}}_{Q}$ and kurtosis $\vec{\boldsymbol{\kappa}}_{Q}$, the measurement noise PDF represented by the covariance ${\boldsymbol{R}}$, skewness $\vec{\boldsymbol{\gamma}}_{R}$ and kurtosis $\vec{\boldsymbol{\kappa}}_{R}$, and the pre-defined $\delta$ value.}
    \KwResult{Updated state PDF for $k\geq 1$, represented by the mean ${\vec{\boldsymbol{x}}}_m(k)$, covariance ${\boldsymbol{P}}_x(k)$, skewness $\vec{\boldsymbol{\gamma}}_x(k)$, and kurtosis $\vec{\boldsymbol{\kappa}}_x(k)$.}
    Initialisation\;
    \While{Time epochs not finished}{
        \hspace*{-0.35cm}Time Update:
        \begin{itemize}
            \item Update the kurtosis component according to Eq.~\ref{eq:upd_kappa} if necessary before generating\\ sigma points for the augmented state in Eq.~\ref{eq:aug_pred_state} according to \textcolor{red}{Eqns.~\ref{eq:sig_house}, \ref{eq:weight_alpha} and \ref{eq:weight_beta}};
            \item Propagate sigma points through the nonlinear state transition function in Eq.~\ref{eq:discrete_time_model};
            \item Generate the predicted PDF of the state via Eqs.~\ref{eq:aug_pred_state}-\ref{eq:pred_kurtosis}.
        \end{itemize}
        Measurement Update: 
        \begin{itemize}
            \item Update the kurtosis component according to Eq.~\ref{eq:upd_kappa} if necessary, before generating\\ sigma points for the augmented state in Eq.~\ref{eq:aug_corr_state} according to \textcolor{red}{Eqns.~\ref{eq:sig_house}, \ref{eq:weight_alpha} and \ref{eq:weight_beta}};
            \item Propagate sigma points through the nonlinear measurement function in Eq.~\ref{eq:meas_system};
            \item Generate the predicted PDF of the measurement via Eqs.~\ref{eq:aug_corr_state}-\ref{eq:corr_kurtosis}.
        \end{itemize}
    }
\end{algorithm*}

\section[SRHOUSE]{New $w$-HOUSE in a Square-Root Form}
\label{sec:srhouse}
\textcolor{red}{As mentioned earlier, the weight $w^{2n+1}$ of the last sigma point can become negative in Eq.~\ref{eq:sig_house}, which can compromise the positive semi-definiteness of the covariance update, leading to failure of HOUSE. To solve this problem, the original $\delta$-HOUSE intentionally increases the smaller values of kurtosis to ensure $w^{2n+1}$ stay nonnegative, according to Eq.~\ref{eq:upd_kappa}. However, this action distorts the actual state distribution as pointed out by the authors in~\cite{stojanovski2021house}.}

Meanwhile, it is proved the square-root formulation can improve the numerical stability and guarantee the positive semi-definiteness of the state covariances~\cite{merwe2001square}. To mitigate the aforementioned issue, this work develops a new variant of HOUSE, dubbed $w$-HOUSE, by utilising the recursive update of square-root form of the covariance matrix rather than the covariance matrix itself. \textcolor{red}{A rank-one Cholesky update or downdate is performed subsequently, depending on the sign of $w^{2n+1}$, which alleviates the constraint of having a nonnegative $w^{2n+1}$ so that avoids distorted higher moments by defining the minimum kurtosis threshold values in the original $\delta$-HOUSE.} This threshold value of kurtosis is still useful when kurtosis of state tends to grow along with time. Like the original $\delta$-HOUSE, \textcolor{red}{a new parameter $w$ is introduced to prevent $w^{2n+1}$ of the last sigma point from growing wildly negative.} More specifically, if $w^{2n+1}$ exceeds the threshold value $w$, it will be reset as zero and other $2n$ weights will also be recalculated. Revisit the calculation of weights derived from the general form of Eq.~\ref{eq:sig_house}:
\begin{gather}
    w^j = \frac{1}{\vec{\boldsymbol{\alpha}}^j(\vec{\boldsymbol{\alpha}}^j + \vec{\boldsymbol{\beta}}^j)},\notag\\
    w^{j+n} = \frac{1}{\vec{\boldsymbol{\beta}}^j(\vec{\boldsymbol{\alpha}}^j + \vec{\boldsymbol{\beta}}^j)},\label{eq:sig_srhouse}\\
    w^{2n+1} = 1 - \sum_{j=1}^{2n}w^j,\notag
\end{gather}
where 
\begin{gather}
    \vec{\boldsymbol{\alpha}}^j=\frac{\vec{\boldsymbol{\gamma}}^j+\sqrt{4\vec{\boldsymbol{\kappa}}^j-3\times(\vec{\boldsymbol{\gamma}}^j\circ\vec{\boldsymbol{\gamma}}^j)}}{2} \label{eq:weight_alpha}
\end{gather}
and
\begin{gather}
    \vec{\boldsymbol{\beta}}^j=\frac{-\vec{\boldsymbol{\gamma}}^j+\sqrt{4\vec{\boldsymbol{\kappa}}^j-3\times(\vec{\boldsymbol{\gamma}}^j\circ\vec{\boldsymbol{\gamma}}^j)}}{2} \label{eq:weight_beta}
\end{gather}
are dependent on the skewness $\vec{\boldsymbol{\gamma}}^j$ and kurtosis $\vec{\boldsymbol{\kappa}}^j$ of the normalised state $\vec{\tilde{\boldsymbol{x}}}^{j}$ in Eqs.~\ref{eq:pred_normalised_state} and \ref{eq:corr_normalised_state} in time update and measurement update, respectively. Note weights $w^j$ and $w^{n+j}$ ($1\leq j\leq n$) are always positive while $w^{2n+1}$ is not. Instead, the proposed $w$-HOUSE does not require $w^{2n+1}$ to be positive as always in the square-root form by using the \textbf{cholupdate} algorithm introduced as below.

If ${\boldsymbol{S}}$ is the original Cholesky factor of ${\boldsymbol{P}}$, then the Cholesky factor of the rank-one update or downdate ${\boldsymbol{P}}\pm\sqrt{\omega}{\boldsymbol{u}}{\boldsymbol{u}}^T$ can be obtained by ${\boldsymbol{S}}=\textbf{cholupdate}{({\boldsymbol{S}},{\boldsymbol{u}},\pm \omega})$. The operation will be executed for $m$ consecutive updates of Cholesky factor if ${\boldsymbol{u}}$ is an $m-$column matrix~\cite{merwe2001square}. An efficient implementation of the \textbf{cholupdate} algorithm is given in~\cite{krause2015more}. 
More specifically, instead of using Eq.~\ref{eq:aug_pred_covariance} in the time update, its original Cholesky factor is used throughout. It is then predicted via the QR decomposition given the deviations of sigma points from the mean and associated weights:\textcolor{red}{
\begin{gather}
    {\boldsymbol{S}}_{a}(k|k-1)=\textbf{qr}\Bigg(\sum_{j=1}^{2(n_x+n_{\omega})} \sqrt{w^{(j)}}\vec{\delta\boldsymbol{x}}_a^{(j)}(k|k-1)\Bigg),
    \label{eq:pred_chol_factor_covariance}
\end{gather}}
followed by the rank-one update or downdate by the ${(2(n_x+n_{\omega})+1)}^{\text{th}}$ sigma point:
\begin{gather}
    {\boldsymbol{S}}_{a}(k|k-1)=\textbf{cholupdate}\Big({\boldsymbol{S}}_{a}(k|k-1),\vec{\delta\boldsymbol{x}}_a^{(1)}(k|k-1),w^{({2(n_x+n_{\omega})+1})}\Big).
    \label{eq:pred_chol_factor_covariance_update}
\end{gather}    
The normalised sigma points are calculated as:
\begin{gather}
    \vec{\tilde{\boldsymbol{x}}}^{(j)}(k|k-1) = [{\boldsymbol{S}}_{x}(k|k-1)]^{-1} \vec{\delta\boldsymbol{x}}^{(j)}(k|k-1),
    \label{eq:pred_chol_factor_normalised_state}
\end{gather}
where $\vec{\delta\boldsymbol{x}}^{(j)}(k|k-1)$ is the first $n_x$ rows of $\vec{\delta\boldsymbol{x}}_{a}^{(j)}(k|k-1)$ and ${\boldsymbol{S}}_{x}(k|k-1)$ is the first $n_x \times n_x$ block of ${\boldsymbol{S}}_{a}(k|k-1)$. Based on thee normalised sigma points, the predicted skewness and kurtosis of the state are calculated. 

In the measurement update, instead of using the state coviarance in Eq.~\ref{eq:aug_corr_covariance}, its original Cholesky factor is used throughout. Likewise, the Cholesky factor of the measurement covariance is calculated via the QR decomposition:\textcolor{red}{
\begin{gather}
    {\boldsymbol{S}}_{z}(k)=\textbf{qr}\Bigg(\sum_{j=1}^{2(n_{x} + n_{\nu})} \sqrt{w^j}\vec{\delta\boldsymbol{z}}^{(j)}(k)\Bigg),
    \label{eq:meas_chol_factor_covariance}
\end{gather}}
followed by the rank-one update or downdate by the ${(2(n_x+n_{\nu})+1)}^{\text{th}}$ sigma point:
\begin{gather}
    {\boldsymbol{S}}_{z}(k)=\textbf{cholupdate}({\boldsymbol{S}}_{z}(k),\vec{\delta\boldsymbol{z}}^{(1)}(k),w^{(2(n_x+n_{\nu})+1)}).
    \label{eq:meas_chol_factor_covariance_update}
\end{gather}
Then the Kalman gain is calculated as:
\begin{gather}
    \boldsymbol{K} = {\boldsymbol{P}}_{xz}(k)/{\boldsymbol{S}}_{z}^T(k)/{\boldsymbol{S}}_{z}(k).
    \label{eq:corr_chol_factor_kalman_gain}
\end{gather}
The Cholesky factor of the covariance of the state is updated as:
\begin{gather}
    {\boldsymbol{S}}_{x}(k) = \textbf{cholupdate}({\boldsymbol{S}}_{x}(k|k-1),\boldsymbol{K}{\boldsymbol{S}}_{z}(k),-1).
    \label{eq:corr_chol_factor_covariance_update}
\end{gather}
Accordingly, the normalised sigma points are calculated as:
\begin{gather}
    \vec{\hat{\boldsymbol{x}}}^{(j)}(k) = [{\boldsymbol{S}}_{x}(k)]^{-1} \vec{\delta\boldsymbol{x}}^{(j)}(k),
    \label{eq:corr_chol_factor_normalised_state}
\end{gather} 
followed by the updated skewness and kurtosis of the state. Accordingly, the new $w$-HOUSE algorithm is summarised in Algorithm~\ref{algo:wHOUSE}. 


\begin{algorithm*}
    \justifying
    \caption{$w$-HOUSE Algorithm}
    \label{algo:wHOUSE}
    \KwData{Initial time epoch, initial state PDF represented by the mean $\vec{\boldsymbol{x}}_m(0)$, Cholesky factorisation of the covariance ${\boldsymbol{S}}_{x}(0)=\textbf{cholesky}({\boldsymbol{P}}_{x}(0))$, the skewness $\vec{\boldsymbol{\gamma}}_{x}(0)$ and kurtosis $\vec{\boldsymbol{\kappa}}_{x}(0)$, the process noise PDF represented by the Cholesky factorisations of the covariance ${\boldsymbol{S}}_{Q}=\textbf{cholesky}({\boldsymbol{Q}})$, skewness $\vec{\boldsymbol{\gamma}}_{Q}$ and kurtosis $\vec{\boldsymbol{\kappa}}_{Q}$, the measurement noise PDF represented by the Cholesky factorisations of the covariance ${\boldsymbol{S}}_{R}=\textbf{cholesky}({\boldsymbol{R}})$, skewness $\vec{\boldsymbol{\gamma}}_{R}$ and kurtosis $\vec{\boldsymbol{\kappa}}_{R}$, and the pre-defined $w$ value.}
    \KwResult{Updated state PDF for $k\geq 1$, represented by the mean ${\vec{\boldsymbol{x}}}_m(k)$, Cholesky
    factorisation of the covariance ${\boldsymbol{P}}_x(k)$, skewness $\vec{\boldsymbol{\gamma}}_x(k)$, and kurtosis $\vec{\boldsymbol{\kappa}}_x(k)$.}
    Initialisation\;
    \While{Time epochs not finished}{
        \hspace*{-0.35cm}Time Update:
        \begin{itemize}
            \item Update the kurtosis component according to Eq.~\ref{eq:upd_kappa} if necessary before generating\\ sigma points for the augmented state in Eq.~\ref{eq:aug_pred_state} according to \textcolor{red}{Eqns.~\ref{eq:sig_house}, \ref{eq:weight_alpha} and \ref{eq:weight_beta}};
            \item Propagate sigma points through the nonlinear state transition function in Eq.~\ref{eq:discrete_time_model};
            \item Generate the predicted PDF similar to $\delta$-HOUSE. Eqs.~\ref{eq:aug_pred_state}, \ref{eq:aug_pred_skewness}-\ref{eq:pred_deviations}, \ref{eq:pred_chol_factor_covariance}-\ref{eq:pred_chol_factor_normalised_state}, and \ref{eq:pred_skewness}-\ref{eq:pred_kurtosis}.
        \end{itemize}
        Measurement Update: 
        \begin{itemize}
            \item Update the kurtosis component according to Eq.~\ref{eq:upd_kappa} if necessary before generating\\ sigma points for the augmented state in Eq.~\ref{eq:aug_corr_state} according to \textcolor{red}{Eqns.~\ref{eq:sig_house}, \ref{eq:weight_alpha} and \ref{eq:weight_beta}};
            \item Propagate sigma points through the nonlinear measurement function in Eq.~\ref{eq:meas_system};
            \item Generate the updated PDF of the state via Eqs.~\ref{eq:aug_corr_state}, \ref{eq:aug_corr_skewness}-\ref{eq:corr_meas_deviations}, \ref{eq:meas_chol_factor_covariance}-\ref{eq:meas_chol_factor_covariance_update}, \ref{eq:corr_cross_covariance}, \ref{eq:corr_chol_factor_kalman_gain}, \ref{eq:corr_mean}, \ref{eq:corr_chol_factor_covariance_update}-\ref{eq:corr_chol_factor_normalised_state}, \\and \ref{eq:corr_skewness}-\ref{eq:corr_kurtosis}.
        \end{itemize}
    }
\end{algorithm*}

\section{Numerical Simulations and Experiments}
To demonstrate the effectiveness of the proposed $w$-HOUSE, numerical simulations and experiments were conducted in this section. The first example was inherited from the projectile example presented in~\cite{stojanovski2021house}. The second example was a synthetic OD scenario with radar measurements. The third example was a real-world OD scenario with optical angles.
\label{sec:simulations}

\subsection{Projectile Example}
As the first example, this work implemented the proposed $w$-HOUSE into the Projectile Example presented in~\cite{stojanovski2021house}. The equations of motion are expressed as:
\begin{equation}
    \begin{aligned}
        \dot{x} &= v_x, \\
        \dot{y} &= v_y, \\
        \dot{z} &= v_z, \\
        \dot{v}_x &= -bv\dot{v}_x + f_x, \\
        \dot{v}_y &= -bv\dot{v}_y + f_y, \\
        \dot{v}_z &= -bv\dot{v}_z + f_z - g, 
    \end{aligned}
\end{equation}
where $b$ is a parameter to calculate the atmospheric drag, which is set as a constant of \SI{0.001}{\per\meter}, $g=\SI{9.807}{\meter\per\second\squared}$ is the Earth gravitational acceleration, $v$ is the speed, and $f_x$, $f_y$ and $f_z$ are other specific accelerations. These specific accelerations are taken as independent variables with mean zeros and standard deviations of $\SI{0.01}{\meter\per\second\squared}$. The initial state is $[1000, 1000, 0, 500, 0, 500]^T$ with the unit of \si{\meter} for position components and \si{\meter\per\second} for velocity components. The associated standard deviation values are \SI{250}{\meter} for position components and \SI{100}{\meter\per\second} for velocity components. Azimuth and elevation angles were simulated at \SI{5}{\hertz} as measurements by the following formulas:
\begin{equation}
    \begin{aligned}
        \alpha_{\text{Az}} = \arctan(y,-x) + \nu_{\alpha},\\
        \epsilon_{\text{El}} = \arctan(z,\sqrt{x^2+y^2}) + \nu_{\epsilon},
    \end{aligned}    
\end{equation}    
where $\nu_{\alpha}$ and $\nu_{\epsilon}$ indicate additive measurement noises, both with mean zeros and standard deviations of \SI{1}{\arcminute}. Gaussian and non-Gaussian Pearson type IV distributions were considered in two cases, respectively. For the latter case, all distributions have a kurtosis of 30.0; the initial state and process noise have a skewness of 1.0 and the measurement noise has a skewness of -1.0. For $\delta$-HOUSE, $\delta$ was set to 0. For $w$-HOUSE, $w$ was set to -0.1.

\subsection{Synthetic OD Example with Radar Measurements}
A numberical example similar to the one presented in~\cite{yun2022kernel} was simulated, in which a RSO in LEO is observed by a ground station at the Earth's north pole. The only difference is the measurements were generated according to the elevation threshold instead of randomly choosing the observational time window. The initial orbital state at \SI{0}{\hour} \SI{0}{\minute} \SI{0}{\second} of January 4, 2010 is given in the ECI coordinate system as below:
\begin{equation*}
    \begin{aligned}
        \vec{\boldsymbol{x}}_0 = [7007.2175, 0, 0, 0, 0.6606, 7.5509],
    \end{aligned}
\end{equation*}
with the units of $\si{\meter}$ and $\si{\meter\per\second}$ for position and velocity components, respectively. A non-Gaussian Pearson Type IV distribution was considered for the initial state. Its covariance matrix is given as: 
\begin{figure*}
    \[
    \centering
    \mathbf{P}_0 = \begin{bmatrix}
        1.481 \times 10^8 & 0 & 0 & 0 & -9.237 \times 10^4 & -5.333 \times 10^4 \\
        0 & 2.885 \times 10^7 & 9.994 \times 10^6 & -3.121 \times 10^4 & 0 & 0 \\
        0 & 9.994 \times 10^6 & 5.770 \times 10^6 & -1.242 \times 10^4 & 0 & 0 \\
        0 & -3.121 \times 10^4 & -1.242 \times 10^4 & 3.687 \times 10^1 & 0 & 0 \\
        -9.237 \times 10^4 & 0 & 0 & 0 & 6.798 \times 10^1 & 3.145 \times 10^1 \\
        -5.333 \times 10^4 & 0 & 0 & 0 & 3.145 \times 10^1 & 3.166 \times 10^1 
    \end{bmatrix}.
    \]
\end{figure*}
The initial skewness and kurtosis were set as -1.6 and 10.0, respectively, for all state variables.
 
\begin{figure}[ht]
    \centering
    \includegraphics[width=8cm]{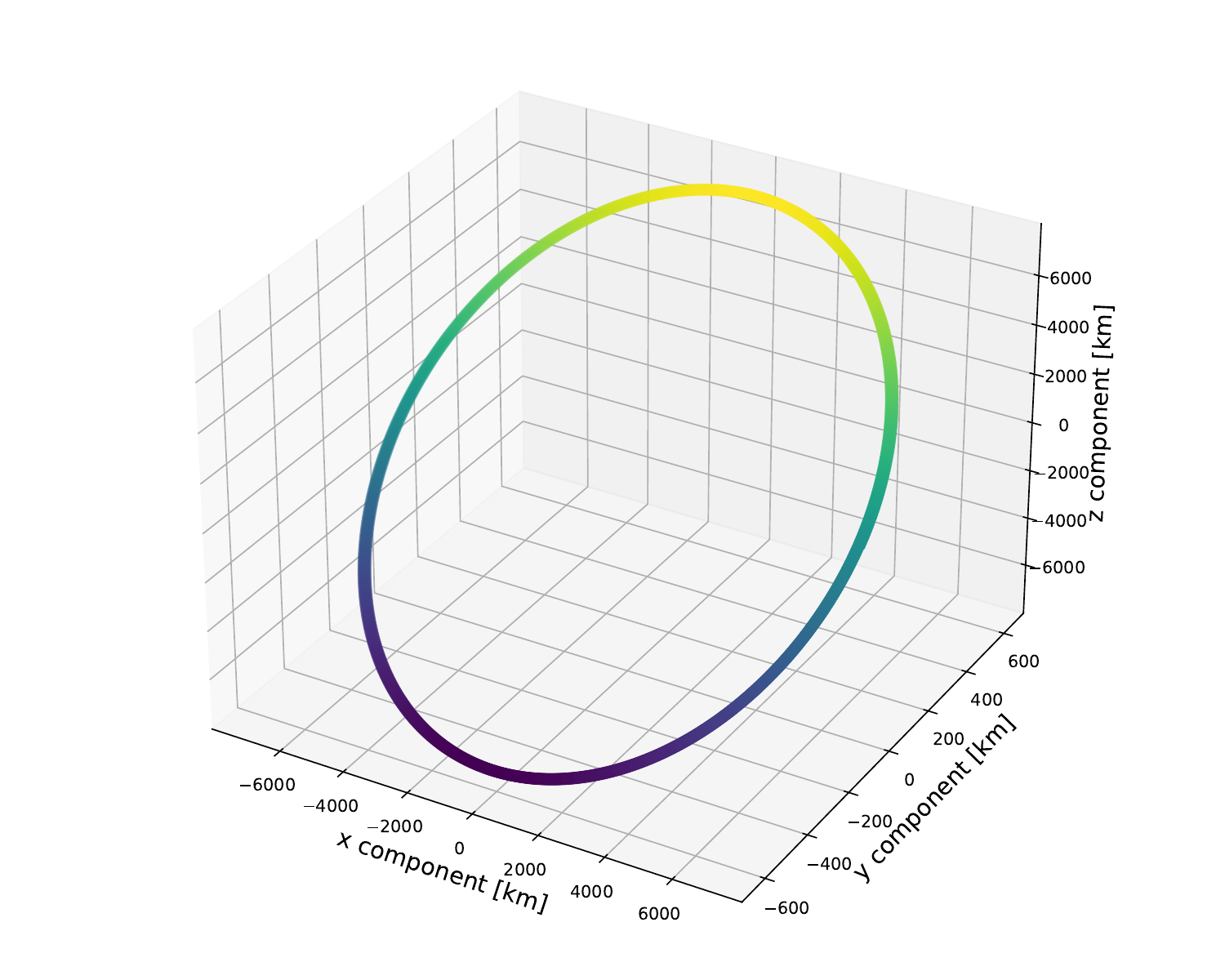}
    \caption{Scatter Plot of the 3D Reference Orbit for the Synthetic OD Example}
    \label{fig:true_traj}
\end{figure}

\begin{figure}[ht]
    \centering
    \includegraphics[width=9cm]{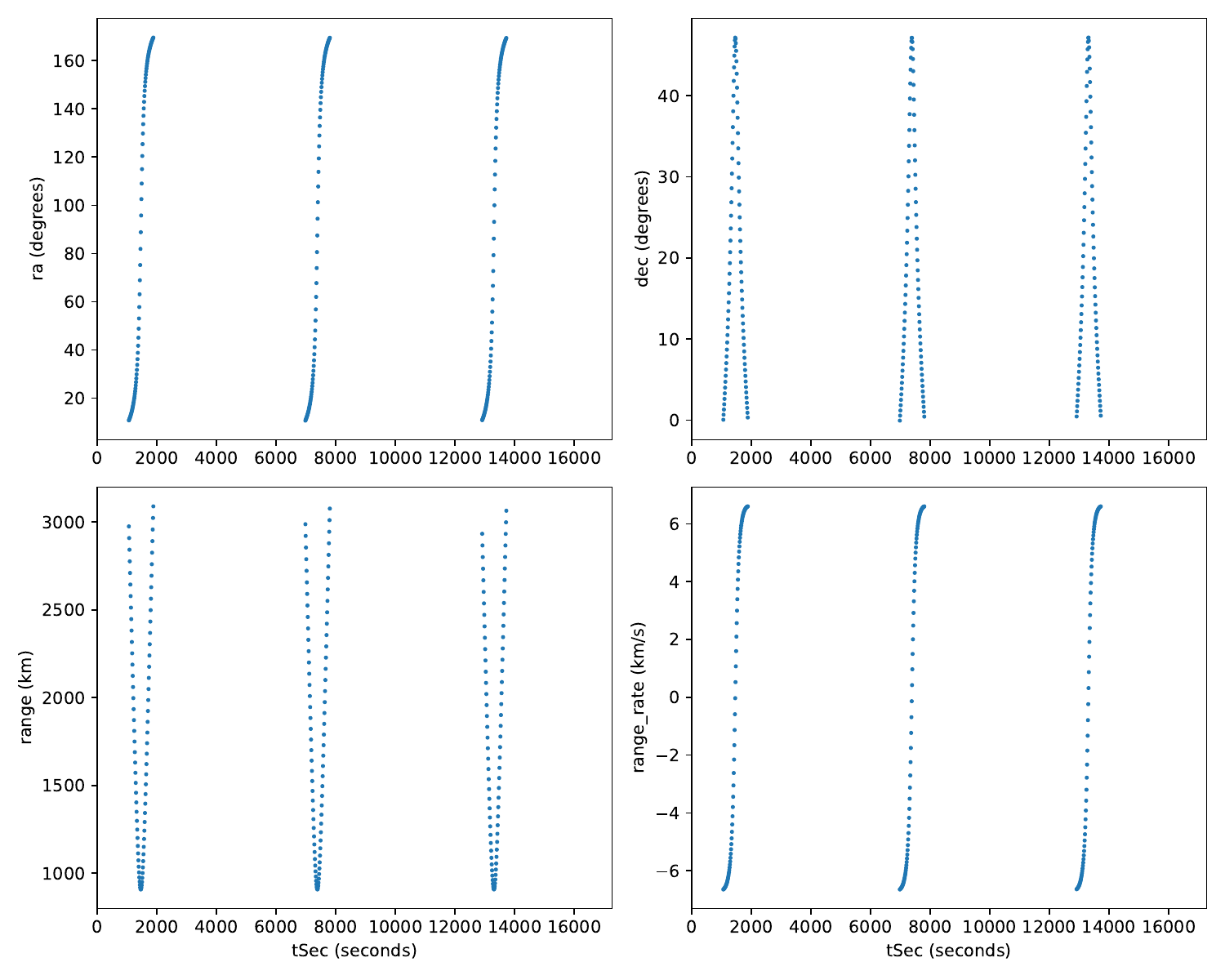}
    \caption{Uncorrupted Measurements for the Synthetic OD Example}
    \label{fig:true_meas}
\end{figure}

As regards the orbital dynamics models for the ground truth, a $40 \times 40$ Earth gravitational potential field was considered using the GGM03S model~\cite{tapley2007ggm03}, together with third-body attractions by Sun and Moon. No surface forces were considered. These orbital dynamics models were used to generate synthetic measurements at an interval of \SI{30}{\second}. The ground truth of the trajectory is shown in Fig.~\ref{fig:true_traj}. 
A radar sensor was considered in this scenario so a set of range, range rate and two angles were simulated when the satellite was above the horizon. A Pearson Type IV distribution was considered, with standard deviations of $\SI{100}{\meter}$, $\SI{0.3}{\meter\per\second}$ and $\SI{100}{\arcsecond}$, and skewness of -1.0 and kurtosis of 30.0 for all measurement variables. The uncorrupted measurements are shown in Fig.~\ref{fig:true_meas}. 

The filter only used a $10 \times 10$ Earth gravity model for orbit prediction. A Gaussian process noise was considered and its covariance is given as a diagonal matrix with the standard deviations of \textcolor{red}{$\SI{2}{\meter}$ for all position components and $\SI{0.01}{\meter\per\second}$ for all velocity components}. For $\delta$-HOUSE, $\delta$ was set to 0. For $w$-HOUSE, $w$ was set to -0.1.

\subsection{Real-World OD Example with Optical Angles}

The Chinese Academy of Sciences has developed and installed a couple of electro-optical arrays at Changchun Observatory that monitor RSOs cross different orbital domains from ground, ranging from LEO to higher orbits. Their tracking facilities are operating in different modes, either to track a mount of specific targets or to survey a complete region~\cite{yang2021single}. In general, an all-sky survey yields short sequences of angular measurements, covering only a small portion of \textcolor{red}{a complete orbital revolution}. Hence, usually it is impossible to retrieve the accurate and reliable orbital state information from a single track. After track-to-track association for different time epochs~\cite{yang2021single}, multiple tracks of right ascension and declination were obtained for the Sentinel 6A satellite in LEO and tested by six filters in the OD process. 

The precise orbit determination (POD) solutions of this satellite are accessible via the European Space Agency Copernicus Open Access Hub~\cite{sanchez2022sentinel}. They were solved via processing measurements from onboard global navigation satellite systems and have reached decimetre-level of accuracy for the three dimensional (3D) position. These POD solutions were interpolated to the optical measurement epochs and used as a benchmark for all filters to compare to. 

The MEE propagator aligned with the intervals of two adjacent measurements for state prediction. The satellite mass was \SI{961.831}{\kilo\gram}. \textcolor{red}{Constant values} of 1.23 and 2.0 were used for the coefficients of solar radiation pressure (SRP) and atmospheric drag calculations, respectively. They are the two main surface forces acting on the LEO satellites. The effective areas were \SI{8.73}{\square \meter} for SRP and \SI{4.09}{\square \meter} for drag. As for field forces, the $100 \times 100$ GRACE Gravity Model GGM03S was considered, together with the third-body attractions from the Sun and Moon whose locations were obtained via the Jet Propulsion Laboratory ephemeris DE440~\cite{park2021jpl}. The solid Earth tide, ocean tide loading and relativistic effect were also considered for as field forces~\cite{yang2022phifa}. The Earth rotational parameters were taken from the Center for Orbit Determination in Europe~\cite{hefty2000analysis} to facilitate the transformations between the ECI frame and the Earth centred Earth fixed (ECEF) frame. It is indicated one-day orbit propagation via high-fidelity models can reach less than \SI{20}{\meter} in 3D position error and less than \SI{5}{\milli\meter\per\second} in 3D velocity error for the Sentinel 6A satellite.  

During the measurement outage, the maximum time step was fixed at \SI{180}{\second} for the propagator to work with. Other time steps can also be chosen, but the process noise covariance matrix need to be adjusted accordingly. 
The initial condition for the satellite orbit can be generated via the method in~\cite{yang2021single} or using the two-line element (TLE). In this work, a set of TLE on the first day of the tracking window has been downloaded from \url{www.spacetrack.org} and propagated to the first observational epoch. The generated initial orbital state in the ECI frame is:
\begin{equation}
    \begin{aligned}
        \vec{\boldsymbol{x}}_{0} &= [-3669.576, 1040.419, 6705.990, \\
        &\phantom{=}\, -5.051, -4.687, -2.036]^T,
        \label{eq:ini_state_cartesian}
    \end{aligned}
\end{equation}  
where the unit is \si{\meter} for position components and \si{\meter\per\second} for velocity components. 
Without knowing the exact statistical information of the TLE data, the initial covariance matrix is assumed as a diagnoal matrix, given by: 
\begin{gather}
    \mathbf{P}_0 = \text{diag}(1.0, 0.01, 0.04, 0.01,0.25,0.04),
    \label{eq:ini_p_cartesian}
\end{gather} 
with the unit of $\si[]{\meter}^2$ for position components and \si[]{\meter\squared\per\second\squared} for velocity components. 
The oscculating MEE for the initial state is obtained by converting the Cartesian coordinate in Eq.~\ref{eq:ini_state_cartesian} into oscculating COE first, followed by the conversion of COE to MEE via Eq.~\ref{eq:mee_from_coe}:
\begin{figure*}
    \begin{equation}
        \vec{\boldsymbol{x}}_{\text{MEE},0} = 
        \begin{bmatrix}
            \num{7.70868e6}, \num{7.28204e{-4}}, \num{-3.56713e{-4}},\num{0.530838},\num{0.376335},\num{8.78592}
        \end{bmatrix}^T,
        \label{eq:ini_state_mee}
    \end{equation}
\end{figure*}
where the unit for the first element is $\si{\meter}$, the unit for the last element is $\si{\radian}$, and the rest elements are unitless. 
The corresponding covariance matrices of the initial state $\mathbf{P}_{\text{MEE},0}$ and covariance of the process noise are obtained via the unscented transformation~\cite{binz2018direct} and given in Eqs.~\ref{eq:ini_p_mee} and \ref{eq:q_mee}, respectively.
\begin{figure*}
    \small
    \centering
    \begin{equation}
        \mathbf{P}_{\text{MEE},0} =
        \begin{bmatrix}
            \num{1.5519e6} & \num{-0.0632625} & \num{0.140824} & \num{-0.0455456} & \num{-0.0209138} & \num{0.0922962} \\
            \num{-0.0632625} & \num{7.58718e-9} & \num{-3.17609e-9} & \num{1.88519e-9} & \num{-1.85111e-9} & \num{-1.71262e-9} \\
            \num{0.140824} & \num{-3.17609e-9} & \num{1.47201e-8} & \num{-4.51409e-9} & \num{-3.20395e-9} & \num{1.00016e-8} \\
            \num{-0.0455456} & \num{1.88519e-9} & \num{-4.51409e-9} & \num{1.82168e-9} & \num{4.914e-10} & \num{-3.24568e-9} \\
            \num{-0.0209138} & \num{-1.85111e-9} & \num{-3.20395e-9} & \num{4.914e-10} & \num{2.11749e-9} & \num{-2.24564e-9} \\
            \num{0.0922962} & \num{-1.71262e-9} & \num{1.00016e-8} & \num{-3.24568e-9} & \num{-2.24564e-9} & \num{7.21086e-9} \\
        \end{bmatrix},
        \label{eq:ini_p_mee}
    \end{equation}
\end{figure*}

\begin{figure*}[!ht]
    \small
    \centering
    \begin{equation}
        \mathbf{Q}_{\text{MEE}} =
        \begin{bmatrix}
            \num{461.406} & \num{-4.79589e-5} & \num{3.56282e-5} & \num{1.86265e-9} & \num{-9.31323e-10} & \num{-2.98023e-8} \\
            \num{-4.79589e-5} & \num{5.67655e-12} & \num{-2.77737e-12} & \num{-2.44054e-16} & \num{1.81495e-16} & \num{2.76949e-15} \\
            \num{3.56282e-5} & \num{-2.77737e-12} & \num{4.00153e-12} & \num{-4.98489e-16} & \num{3.7058e-16} & \num{3.90877e-15} \\
            \num{1.86265e-9} & \num{-2.44054e-16} & \num{-4.98489e-16} & \num{6.33216e-13} & \num{-4.68597e-13} & \num{-6.82121e-13} \\
            \num{-9.31323e-10} & \num{1.81495e-16} & \num{3.70635e-16} & \num{-4.68625e-13} & \num{3.50248e-13} & \num{5.09814e-13} \\
            \num{1.49012e-8} & \num{2.76949e-15} & \num{3.9092e-15} & \num{-6.82121e-13} & \num{5.0937e-13} & \num{7.67386e-13} \\
        \end{bmatrix}
        \label{eq:q_mee}
    \end{equation}
\end{figure*}
An initial skewness of -1.6 and an initial kurtosis of 15.0 are given for each of six MEEs.  

\begin{figure*}[!ht]
    \centering
    \includegraphics[width=16cm]{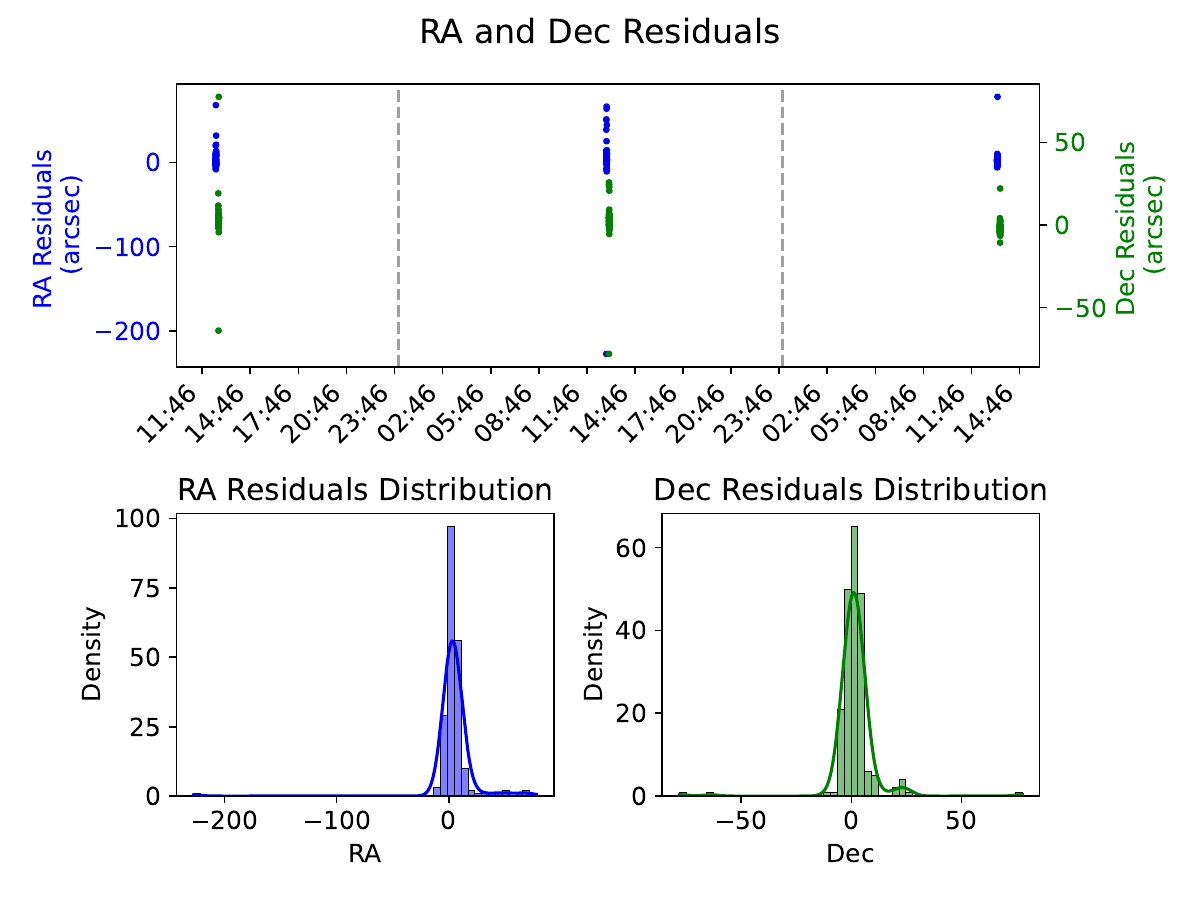}
    \caption{Pre-residuals of Right Ascension and Declination Measurements Based on Precise Orbit Ephemeris of Sentinel 6A Satellite}
    \label{fig:ccdata_pre_res}
\end{figure*}

The coordinate of the ground station in the ECEF frame is: 
\begin{gather}
    \vec{\boldsymbol{x}}_{\text{stn}} = [-2730.0, 3714.0, 4394.0, 0, 0, 0]^T, 
\end{gather} 
where the unit is \si[]{\kilo\meter} for position components and \si[]{\meter\per\second} for velocity components. 
During the three-day tracking campaign starting on 15/05/2021, the Sentinel 6A satellite was only observed by three short arcs, each spanning approximately \SI{2}{\minute}. The pre-residuals of the right ascension and declination angles were calculated based on POD solutions of Sentinel 6A mentioned above, shown in Fig.~\ref{fig:ccdata_pre_res}. Note that the declination residuals were shiftted by \SI{30}{\minute} intentionally for better visualisation. It is indicated that both angles are subject to some errors and large outliers, which lead to non-Gaussian phenomenon of these measurements shown by the distribution plots in the bottom subfigures of Fig.~\ref{fig:ccdata_pre_res}. The statistics of two angular residuals is summarised in Table \ref{tab:ccdata_statistics}.

\begin{table*}[htbp]
    \centering
    \caption{Statistics of Right Ascension and Declination Pre-residuals}
    \label{tab:ccdata_statistics}
    \begin{tabular}{lcccccc}
      \toprule
      & \textbf{Mean (\si{\arcsecond})} & \textbf{Std. Dev. (\si{\arcsecond})} & \textbf{RMS (\si{\arcsecond})} & \textbf{Skewness} & \textbf{Kurtosis} \\
      \midrule
      \textbf{Right Ascension } & \SI{4.650}{} & \SI{20.141}{} & \SI{20.671}{} & -6.528 & 84.349 \\
      \textbf{Declination } & \SI{1.460}{} & \SI{10.319}{} & \SI{10.421}{} & -1.196 & 36.684 \\
      \bottomrule
    \end{tabular}
\end{table*}

\section{Results and Discussions}
\label{sec:results}
\subsection{Projectile Example}
\subsubsection{Accuracy}
\begin{figure*}[ht]
    \centering
    \includegraphics[width=12cm]{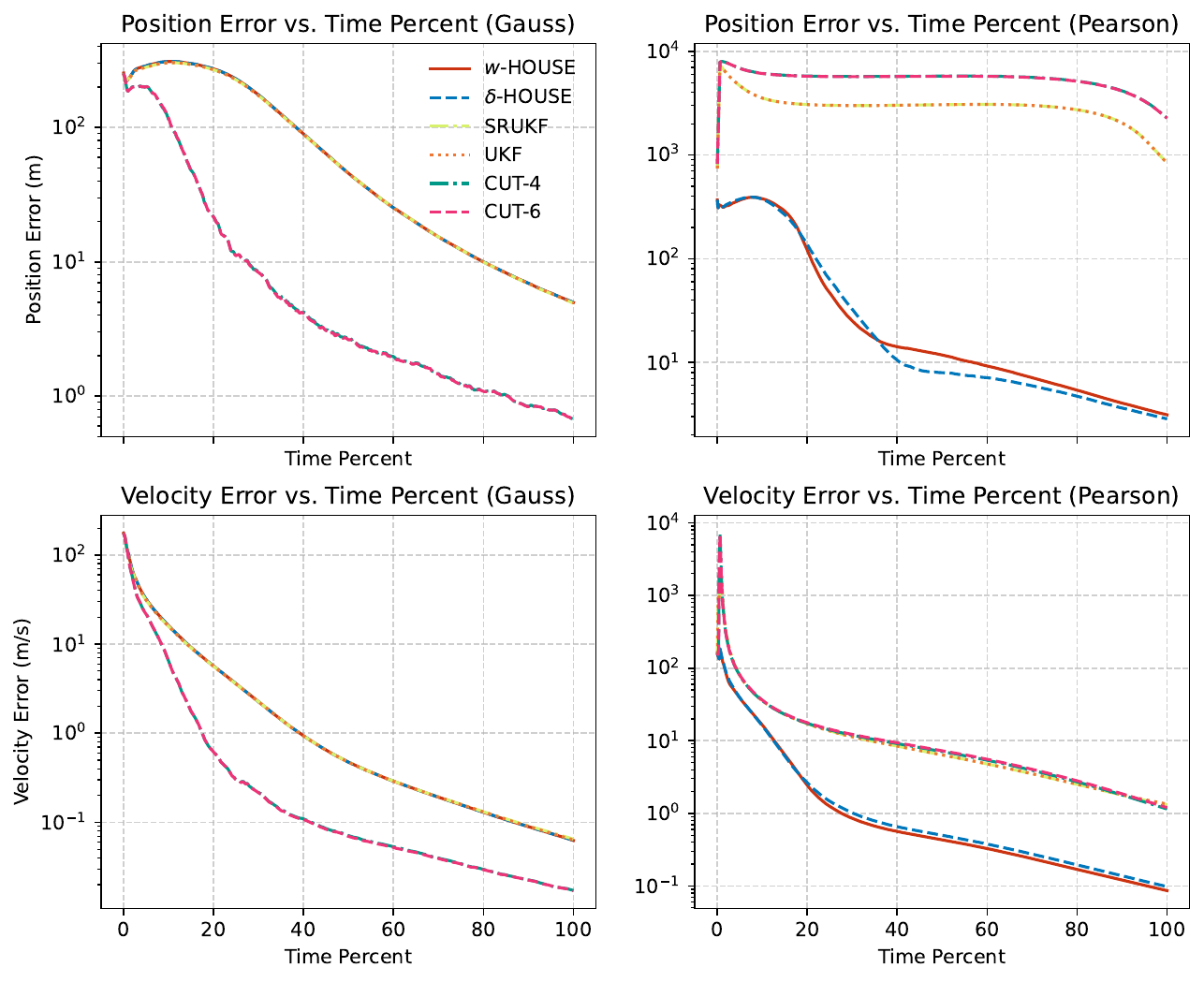}
    \caption{Estimation Errors for the Projectile Example}
    \label{fig:proj_err}
\end{figure*}

100 Monte Carlo simulations (MCS) have been implemented by six filters, including $w$-HOUSE, $\delta$-HOUSE, SRUKF, UKF, CUT-4 and CUT-6. Estimation errors are shown in Fig.~\ref{fig:proj_err}. The upper subfigures indicate the 3D position errors while the lower subfigures indicate the 3D velocity errors. Note the $96^{\text{th}}$ trial has been excluded for generating these plots due to large errors for the first four filters. For the Gaussian case in the left column, CUT-4 and CUT-6 result in both smaller position and velocity errors compared to other four filters including $w$-HOUSE, generating two groups of curves. For the non-Gaussian Pearson type IV case, two HOUSE filters stay close to each other and generate more accurate estimates compared to other filters. In particular, the errors quickly decrease by both HOUSE filters. In two cases, SRUKF overlaps with UKF for all types of errors, and $w$-HOUSE yields very similiar results to $\delta$-HOUSE.  

\begin{figure*}[ht]
    \centering
    \includegraphics[width=12cm]{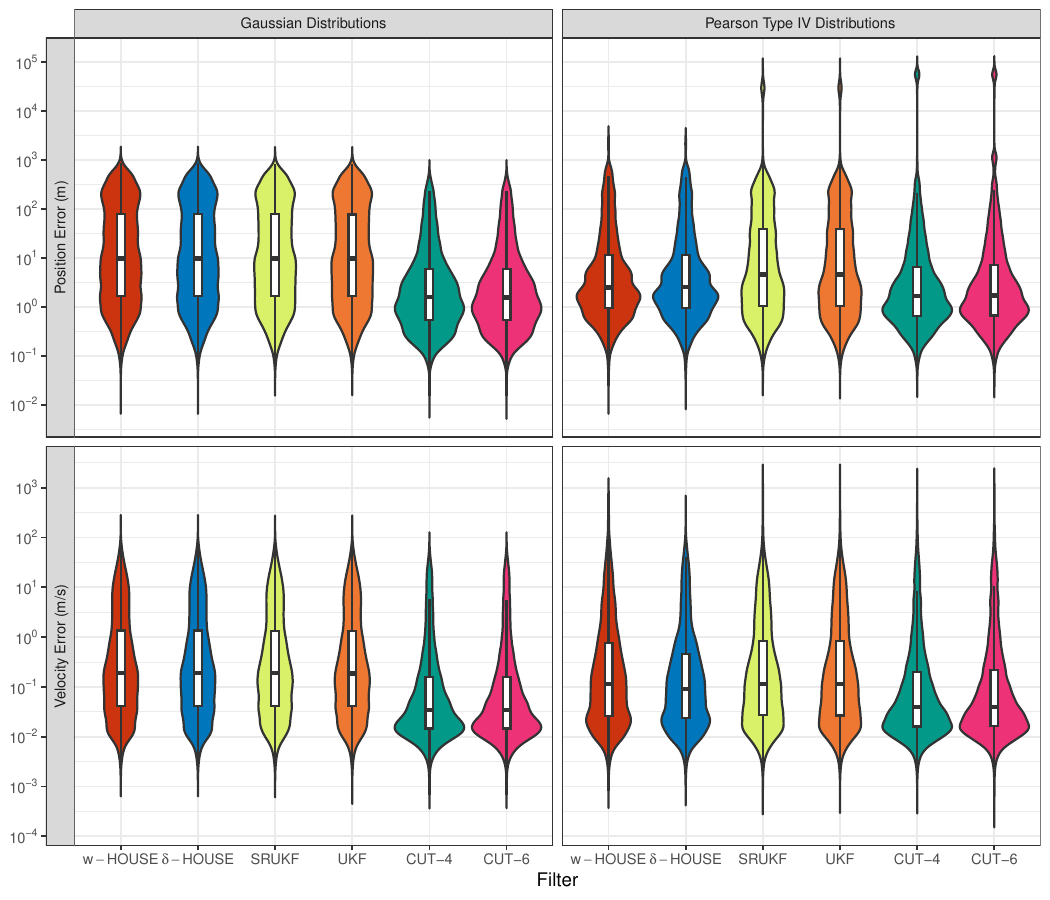}
    \caption{Distribution of Estimation Error for the Projectile Example}
    \label{fig:proj_rmse}
\end{figure*}

\textcolor{red}{Fig.~\ref{fig:proj_rmse} shows the distribution of estimation errors for different filters. The box represents the interquartile range (the range between Q1 and Q3), with the line inside representing the median and the whiskers representing the range of the data. Outliers appear clearly in the right top subfigure for the position errors using the two UKFs and two CUT filters for the Pearson type IV distribution. This is because the Pearson type IV distribution has larger kurtosis values than the Gaussian distribution. As a result, these results prove that both $w$-HOUSE and $\delta$-HOUSE, which directly account for the fourth-order moment of kurtosis, are more robust in the presence of outliers than two UKFs and two CUT filters.}

\subsubsection{Consumption Time}
\begin{figure}[ht]
    \centering
    \includegraphics[width=8cm]{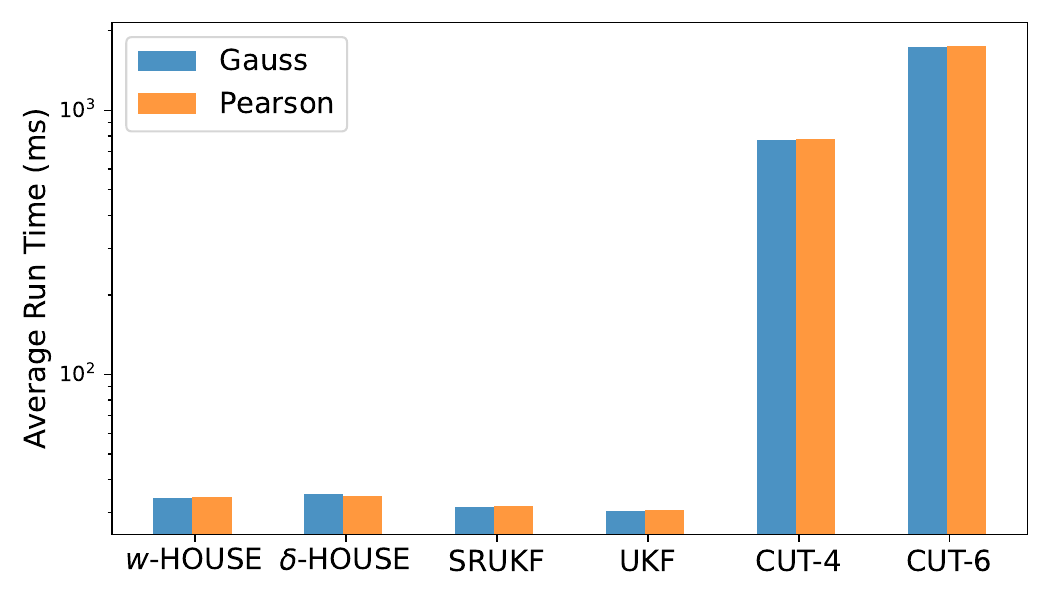}
    \caption{Runtime for the Projectile Example}
    \label{fig:proj_runtime}
\end{figure}

The average runtime over 100 MCS for all six filters are compared in Fig.~\ref{fig:proj_runtime}. Two HOUSE filters take slightly more computational time than two UKFs. The runtime for CUT-4 and CUT-6 are significantly larger. The $w$-HOUSE is even slightly more efficient than $\delta$-HOUSE. \textcolor{red}{All three figures, including Figs.~\ref{fig:proj_err}-\ref{fig:proj_runtime} present consistent results with the original paper\,\cite{stojanovski2021house}.}

\subsection{Synthetic OD Example with Radar Measurements}
\subsubsection{Accuracy}
\begin{figure*}[!ht]
    \centering
    \includegraphics[width=16cm]{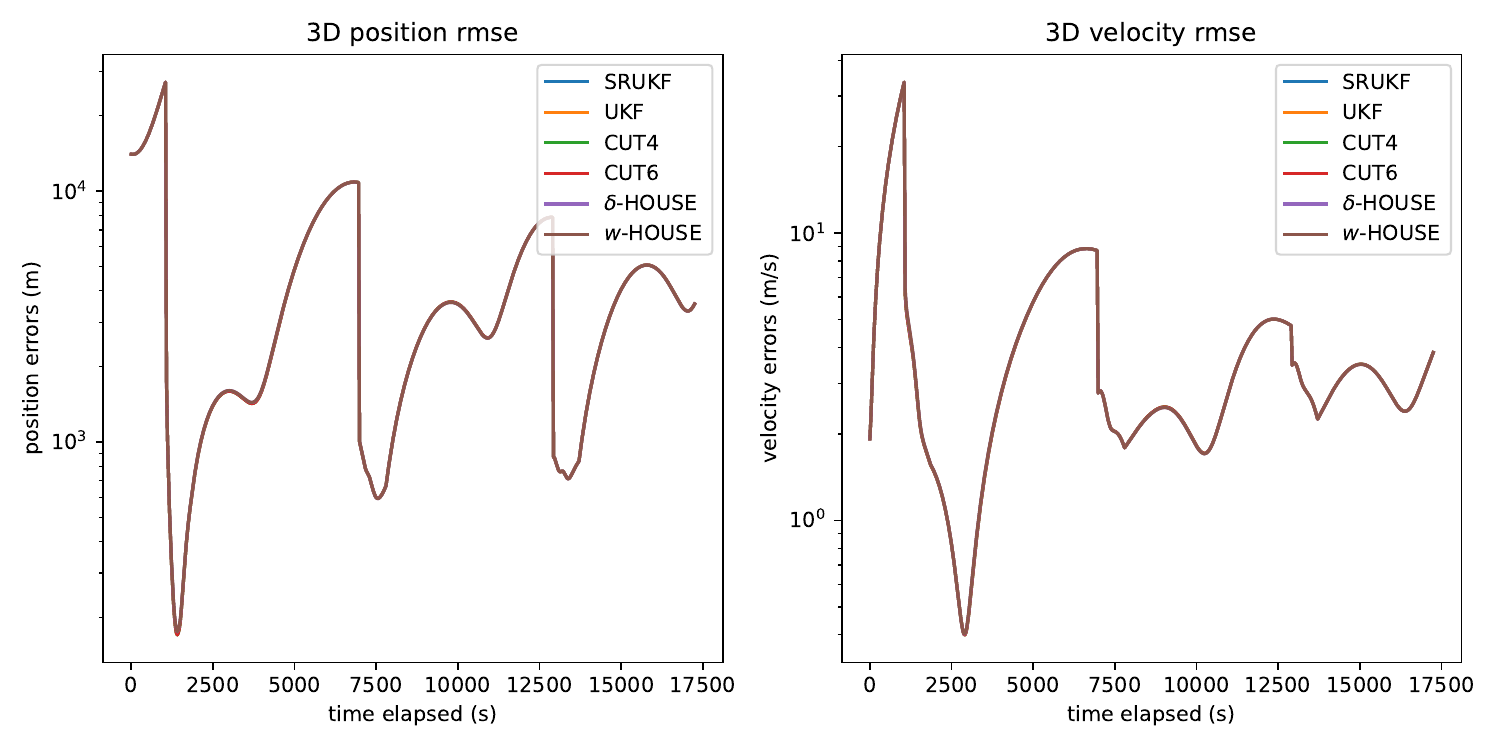}
    \caption{3D Position and Velocity RMSE for All Filters}
    \label{fig:3d_err_synthetic_od}
\end{figure*}

Averaged over 100 Monte Carlo trials, the 3D position and velocity RMSE are depicted in Fig.~\ref{fig:3d_err_synthetic_od} for the synthetic OD example. All filters consistently produce converged position and velocity estimates, with 3D RMSE decreasing from the first track to the last one. Notably, the error curves for all filters substantially overlap, with no discernible differences in both 3D position and velocity errors. Table~\ref{tab:average_error_synthetic_od} presents the average RMSE across all time epochs. The average 3D position RMSE for all filters fall within the range of \SIrange{4421}{4422}{\meter}, while the average 3D velocity RMSE range from \SIrange{4.219}{4.223}{\meter\per\second}. It is worth mentioning that the CUT filters exhibit the largest average 3D position RMSE, whereas $\delta$-HOUSE yields the smallest value. The $\delta$-HOUSE filter does exhibit slightly larger errors compared to the SRUKF and UKF. These results can be attributed to the radar measurements used in this numerical example, which encompass range, range rate, and two angles. The information provided by these measurements effectively mitigates the non-Gaussian phenomena, resulting in similar outcomes for all filters. The next subsection presents a more challenging scenario where only two angles are available.

\begin{table*}[h]
    \centering
    \caption{Average 3D Position and 3D Velocity Errors for All Filters (Units: \si{\meter} and \si{\meter\per\second})}
    \begin{tabular}{cccc}
    \hline
    \textbf{Filter Type} & \textbf{Average 3D Position Error} & \textbf{Average 3D Velocity Error} \\
    \hline
    SRUKF & 4422.117 & 4.219 \\
    UKF & 4422.117 & 4.219 \\
    CUT-4 & 4422.916 & 4.220 \\
    CUT-6 & 4422.812 & 4.220 \\
    $\delta$-HOUSE & 4421.885 & 4.222 \\
    $w$-HOUSE & 4422.313 & 4.223 \\
    \hline
    \end{tabular}
    \label{tab:average_error_synthetic_od}
\end{table*}

\subsubsection{Time Consumption}
\begin{figure}[!ht]
    \centering
    \includegraphics[width=8cm]{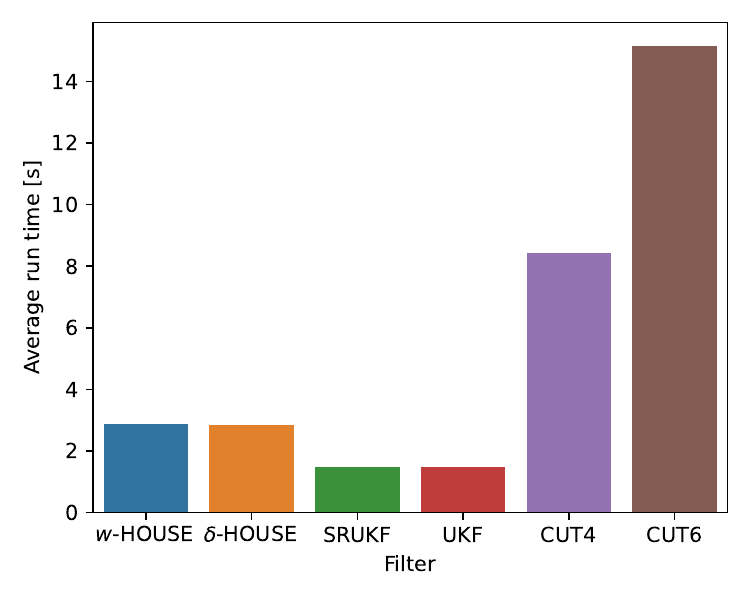}
    \caption{Runtime of Synthetic OD for All Filters}
    \label{fig:run_times_synthetic_od}
\end{figure}

Fig.~\ref{fig:run_times_synthetic_od} depicts the average run times for all filters across 100 MCS. Notably, the SRUKF and UKF filters demonstrate the highest efficiency, while the CUT filters are the most time-consuming. The $w$-HOUSE and $\delta$-HOUSE filters exhibit comparable time consumption, with their runtime being approximately twice that of two UKF filters.
\subsection{Real-World OD Example with Optical Angles}
The OD accuracy and precision via six filters are presented and analysed in the following section. Filters were run with MEE. 
\subsubsection{Absolute Position and Velocity Errors}
\begin{figure*}[!ht]
    \centering
    \includegraphics[width=16.5cm]{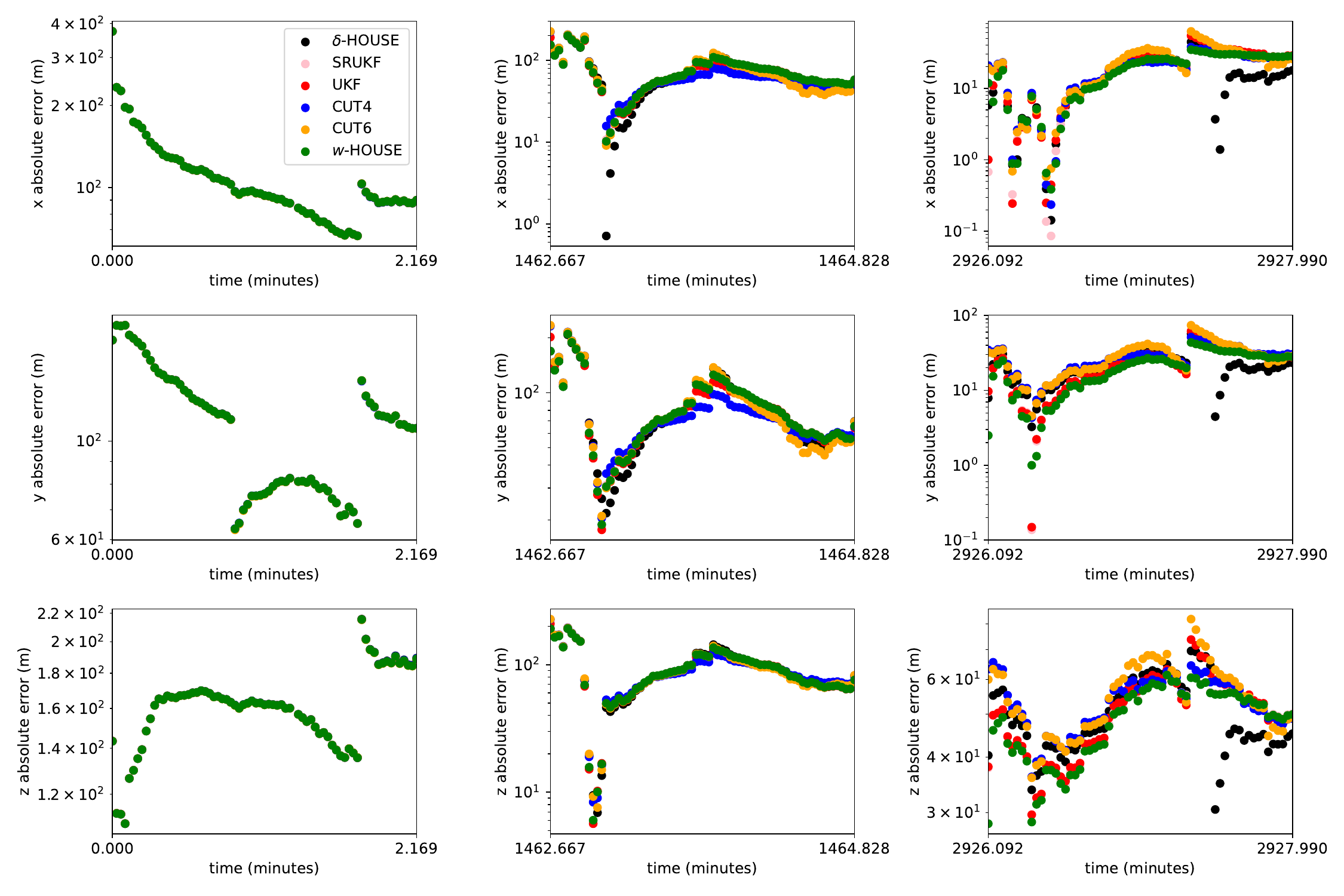}
    \caption{Absolute Position Errors for All Filters}
    \label{fig:all_pos_abs_err}
\end{figure*}

\begin{figure*}[!ht]
    \centering
    \includegraphics[width=16.5cm]{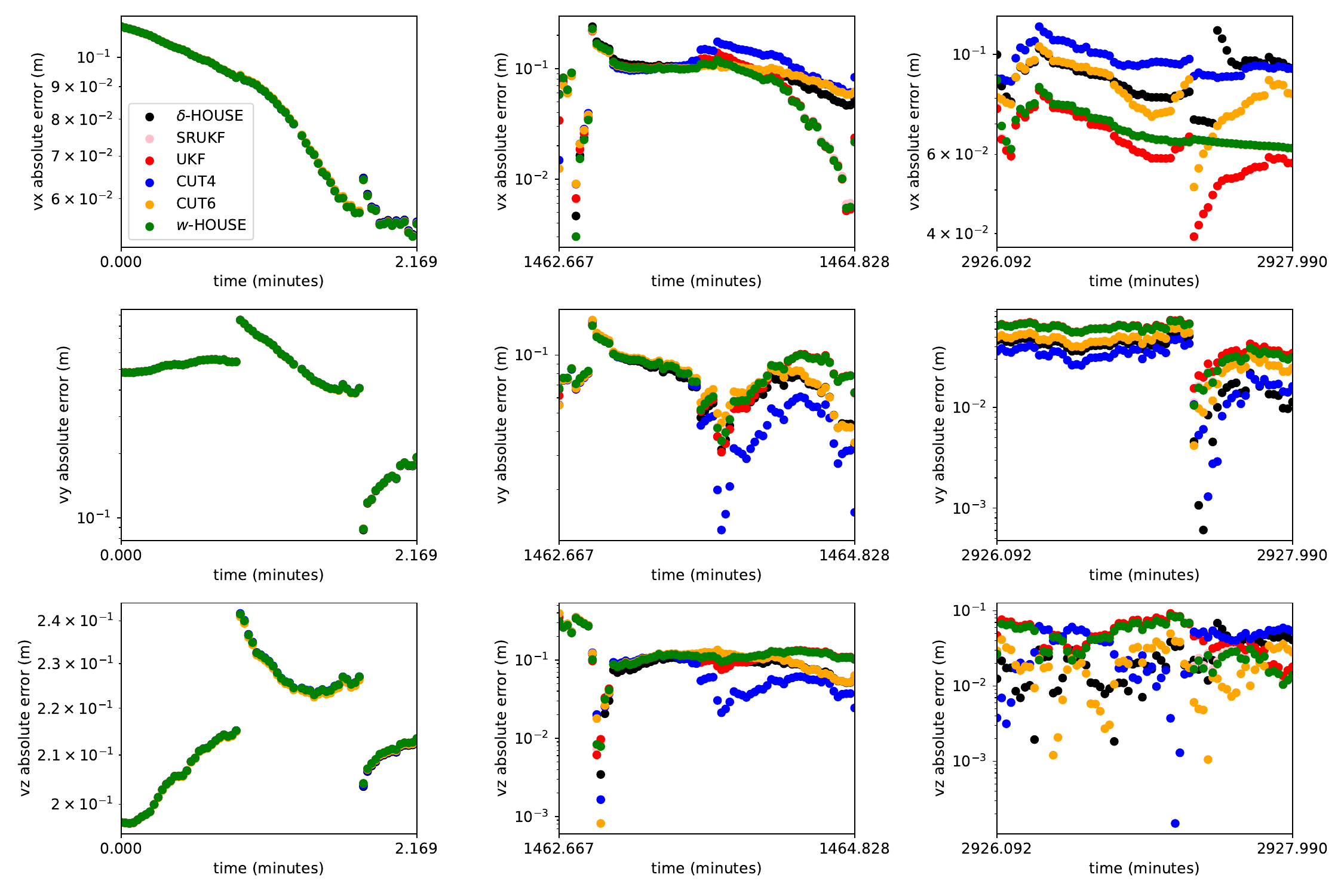}
    \caption{Absolute Velocity Errors for All Filters}
    \label{fig:all_vel_abs_err}
\end{figure*}

The absolute position and velocity errors for three coordinates are shown in Fig.~\ref{fig:all_pos_abs_err} and in Fig.~\ref{fig:all_vel_abs_err}, respectively, each row of subfigures illustrating each component. Three tracking arcs are shown separately in three columns. In each subfigure, absolute errors generated by six filters are illustrated together. Position and velocity errors decrease from the first track to the last. For the first track, the absolute position and velocity errors by all filters stay very close. For the last track, different filters yield larger discrepancies, particularly for velocity estimates. At the beginning phases of tracks 2 and 3 when measurements become available again, all filters are subject to large fluctuations. $\delta$-HOUSE yields some obvious and steep decreases on all position components in the last track. The velocity estimates generated by CUT-4 and CUT-6 scatter around. Given the uncertainty associated with the orbital state does not match well with a Gaussian or uniform PDF, CUT filters present even worse results compared to lower-order UKFs, i.e., the normal UKF and SRUKF. Overall the proposed $w$-HOUSE method presents more smooth position and velocity estimates than all other filters. 

\subsubsection{RMSE Statistics}
The RMSE for three positon components and three velocity components are shown in Table~\ref{tab:pos_rmse_mee_filter} and Table~\ref{tab:vel_rmse_mee_filter}, respectively. All filters yield converged position and velocity estimates in terms of each component and 3D values. The 3D position RMSE decreases from the first track at around \SI{232.5}{\meter} to the last track between \SIrange{58.8}{70.1}{\meter}. The 3D velocity RMSE decreases from the first track at around \SI{0.5}{\meter\per\second} to the last track at around \SI{0.1}{\meter\per\second}. Although the velocity RMSE values are very close across all filters, \textcolor{red}{the $w$-HOUSE stands out as the best among six filters including $\delta$-HOUSE, in terms of 3D position RMSE as shown in Table III. Specifically, the $w$-HOUSE generates a 3D position RMSE that is more than \SI{1}{\meter} smaller compared to $\delta$-HOUSE.}


\begin{table}[htbp]
    \centering
    \caption{Position RMSE for All Filters (Unit: \si{\meter})}
    \label{tab:pos_rmse_mee_filter}
    \begin{tabular}{
        l
        c
        S[table-format=3.3]
        S[table-format=3.3]
        S[table-format=4.3]
        S[table-format=4.3]
    }
        \toprule
        \textbf{Filter Type} & \textbf{Track} & $\mathbf{x}$ & $\mathbf{y}$ & $\mathbf{z}$ & $\textbf{3D Pos}$ \\
        \midrule
        \multirow{3}{*}{$w$-HOUSE} & 1 & 120.730 & 114.857 & 162.229 & 232.564 \\
         & 2 & 83.648 & 91.689 & 98.958 & 158.734 \\
         & 3 & 21.633 & 23.765 & 49.282 & 58.834 \\
        \midrule
        \multirow{3}{*}{$\delta$-HOUSE} & 1 & 120.708 & 114.861 & 162.342 & 232.634 \\
         & 2 & 82.335 & 91.120 & 99.745 & 158.212 \\
         & 3 & 19.492 & 25.461 & 50.657 & 59.953 \\
        \midrule
        \multirow{3}{*}{SRUKF} & 1 & 120.745 & 114.853 & 162.238 & 232.576 \\
         & 2 & 85.164 & 92.946 & 99.336 & 160.498 \\
         & 3 & 21.763 & 24.071 & 49.897 & 59.521 \\
        \midrule
        \multirow{3}{*}{UKF} & 1 & 120.732 & 114.840 & 162.223 & 232.552 \\
         & 2 & 83.209 & 91.185 & 98.620 & 158.001 \\
         & 3 & 25.404 & 27.170 & 51.579 & 63.592 \\
        \midrule
        \multirow{3}{*}{CUT-4} & 1 & 120.714 & 114.821 & 162.271 & 232.568 \\
         & 2 & 79.955 & 88.787 & 98.543 & 154.876 \\
         & 3 & 22.397 & 29.474 & 53.930 & 65.412 \\
        \midrule
        \multirow{3}{*}{CUT-6} & 1 & 120.695 & 114.791 & 162.160 & 232.465 \\
         & 2 & 85.800 & 93.868 & 99.644 & 161.560 \\
         & 3 & 26.641 & 32.489 & 56.165 & 70.141 \\
        \bottomrule
    \end{tabular}
\end{table}

\begin{table}[ht]
    \centering
    \caption{Velocity RMSE for All Filters (Unit: \si{\meter\per\second})}
    \begin{tabular}{
        l
        c
        S[table-format=1.3]
        S[table-format=1.3]
        S[table-format=1.3]
        S[table-format=1.3]
    }
        \toprule
        \textbf{Filter Type} & \textbf{Track} & $\mathbf{vx}$ & $\mathbf{vy}$ & $\mathbf{vz}$ & $\mathbf{3D Vel}$ \\
        \midrule
        \multirow{3}{*}{$w$-HOUSE} & 1 & 0.085 & 0.495 & 0.216 & 0.547 \\
         & 2 & 0.091 & 0.085 & 0.140 & 0.188 \\
         & 3 & 0.069 & 0.054 & 0.047 & 0.099 \\
        \midrule
        \multirow{3}{*}{$\delta$-HOUSE} & 1 & 0.085 & 0.495 & 0.216 & 0.546 \\
         & 2 & 0.099 & 0.076 & 0.127 & 0.178 \\
         & 3 & 0.090 & 0.037 & 0.028 & 0.101 \\
        \midrule
        \multirow{3}{*}{SRUKF} & 1 & 0.085 & 0.495 & 0.216 & 0.547 \\
         & 2 & 0.092 & 0.085 & 0.144 & 0.191 \\
         & 3 & 0.068 & 0.055 & 0.050 & 0.101 \\
        \midrule
        \multirow{3}{*}{UKF} & 1 & 0.085 & 0.495 & 0.216 & 0.547 \\
         & 2 & 0.095 & 0.084 & 0.140 & 0.189 \\
         & 3 & 0.064 & 0.055 & 0.053 & 0.100 \\
        \midrule
        \multirow{3}{*}{CUT-4} & 1 & 0.085 & 0.495 & 0.216 & 0.547 \\
         & 2 & 0.114 & 0.071 & 0.123 & 0.182 \\
         & 3 & 0.097 & 0.030 & 0.040 & 0.109 \\
        \midrule
        \multirow{3}{*}{CUT-6} & 1 & 0.085 & 0.495 & 0.216 & 0.546 \\
         & 2 & 0.097 & 0.081 & 0.139 & 0.188 \\
         & 3 & 0.083 & 0.042 & 0.023 & 0.096 \\
        \bottomrule
    \end{tabular}
    \label{tab:vel_rmse_mee_filter}
\end{table}

\subsubsection{Post-residuals}
\begin{figure*}[!ht]
    \centering
    \includegraphics[width=18cm]{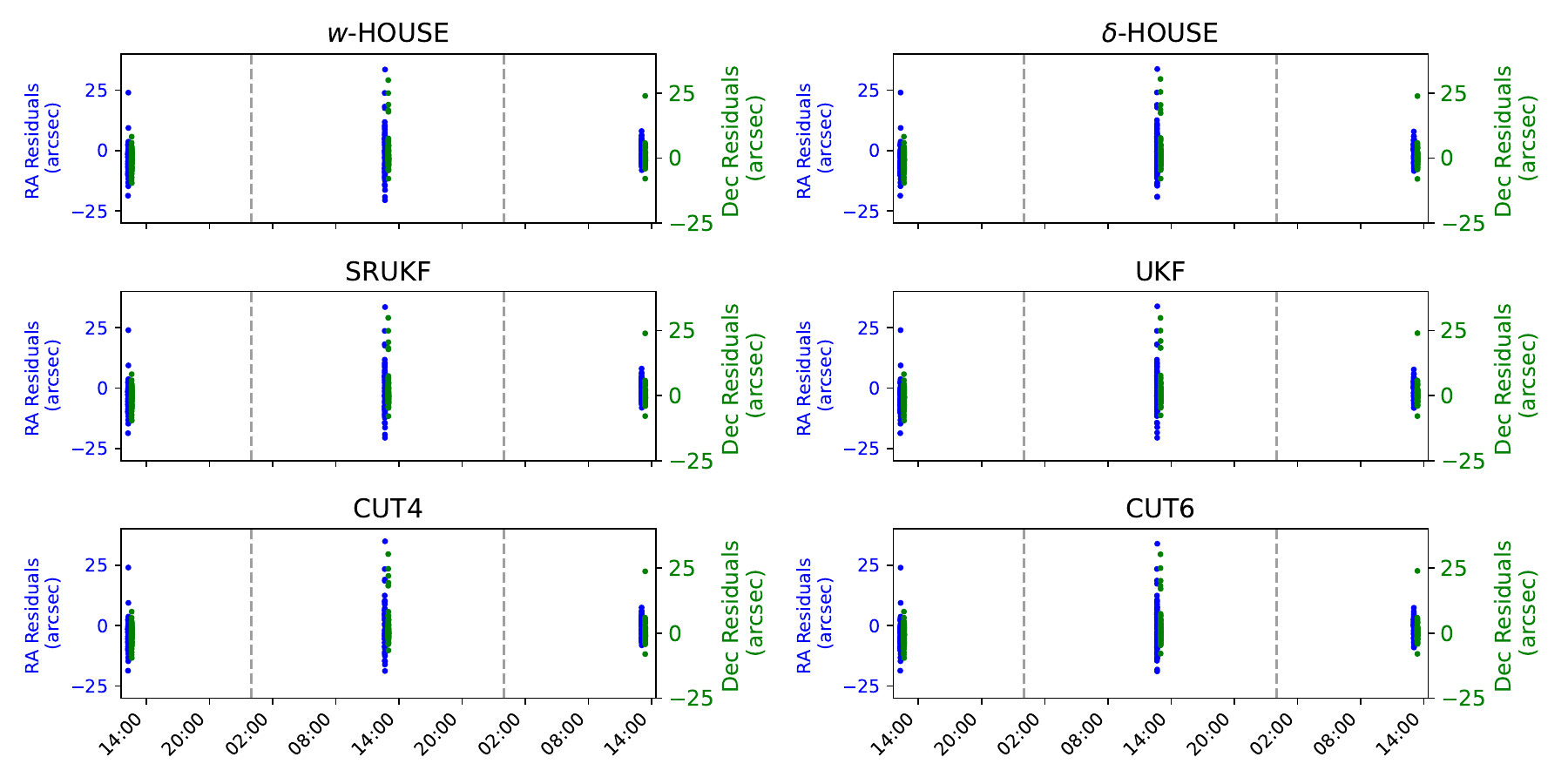}
    \caption{Post-residuals of Two Angles for All Filters}
    \label{fig:all_post_res}
\end{figure*}

The post-residuals of two angles, i.e., right ascension and declination, based on the position estimates were also calculated and they are presented in Fig.~\ref{fig:all_post_res}. Interestingly, the post-residuals for the second track are larger than those for the first and third tracks, which is also indicated by the root mean square (RMS) residuals of two angels shown in Table~\ref{tab:ccdata_rms_post_res}. This is due to the fact that right ascension angles of the second track are more contaminated by the systematic bias indicated by the pre-residual analysis in Fig.~\ref{fig:ccdata_pre_res}. Smallest post-residuals are obtained for the third track, the RMS values of the right ascension are still larger than \SI{9}{\arcsecond} for all filters, while the RMS values of the declination drop down to less than \SI{4}{\arcsecond}. The $w$-HOUSE, $\delta$-HOUSE, SRUKF, and UKF show consistent performance in terms of RMS residuals across all tracks for both right ascension and declination angles. The CUT-4 filter demonstrates slightly higher RMS values while the CUT-6 filter stands out with lower RMS values compared to the other filters, indicating its highest precision. Overall the differences are not significant among all filters. 

\begin{table}[htbp]
    \centering
    \caption{RMS Residuals for All Filters (Unit: \si{\arcsecond})}
    \label{tab:ccdata_rms_post_res}
    \begin{tabular}{ccccc}
    \toprule
    \textbf{Filter Type} & \textbf{Track} & \textbf{RA} & \textbf{Dec} \\
    \midrule
    \multirow{3}{*}{$w$-HOUSE} & 1 & 10.038 & 11.895 \\
    & 2 & 29.605 & 10.631 \\
    & 3 & 9.738 & 3.868 \\
    \midrule
    \multirow{3}{*}{$\delta$-HOUSE} & 1 & 10.038 & 11.896 \\
    & 2 & 29.458 & 10.613 \\
    & 3 & 9.523 & 3.865 \\
    \midrule
    \multirow{3}{*}{SRUKF} & 1 & 10.039 & 11.895 \\
    & 2 & 29.615 & 10.624 \\
    & 3 & 9.726 & 3.870 \\
    \midrule
    \multirow{3}{*}{UKF} & 1 & 10.038 & 11.894 \\
    & 2 & 29.671 & 10.656 \\
    & 3 & 9.461 & 3.876 \\
    \midrule
    \multirow{3}{*}{CUT-4} & 1 & 10.039 & 11.897 \\
    & 2 & 29.499 & 10.601 \\
    & 3 & 9.642 & 3.856 \\
    \midrule
    \multirow{3}{*}{CUT-6} & 1 & 10.037 & 11.893 \\
    & 2 & 29.345 & 10.539 \\
    & 3 & 9.294 & 3.867 \\
    \bottomrule
    \end{tabular}
\end{table}

\subsubsection{Paramertic Analysis of HOUSE Filters} 
\textcolor{red}{The calculation of $w_{2n+1}$ of sigma points in Eq.~\ref{eq:sig_house} is associated with a specific parameter for both $\delta$-HOUSE and $w$-HOUSE. A total of 100 MCS were conducted to analyse the performances of these two filters with varying parameter values. For $\delta$-HOUSE, which requires nonnegative $\delta$ values as per Eq.~\ref{eq:weight_last}, 100 values were selected uniformly from the range of $[0, 0.2]$. Conversely, for $w$ in $w$-HOUSE, where the last weight's nonnegativity is not strictly necesssary (according to Eqns.~\ref{eq:pred_chol_factor_covariance} and \ref{eq:meas_chol_factor_covariance}), 100 values were selected uniformly from the range of $[-0.1, 0.1]$.}

\begin{figure}[!ht]
    \centering
    \includegraphics[width=8cm]{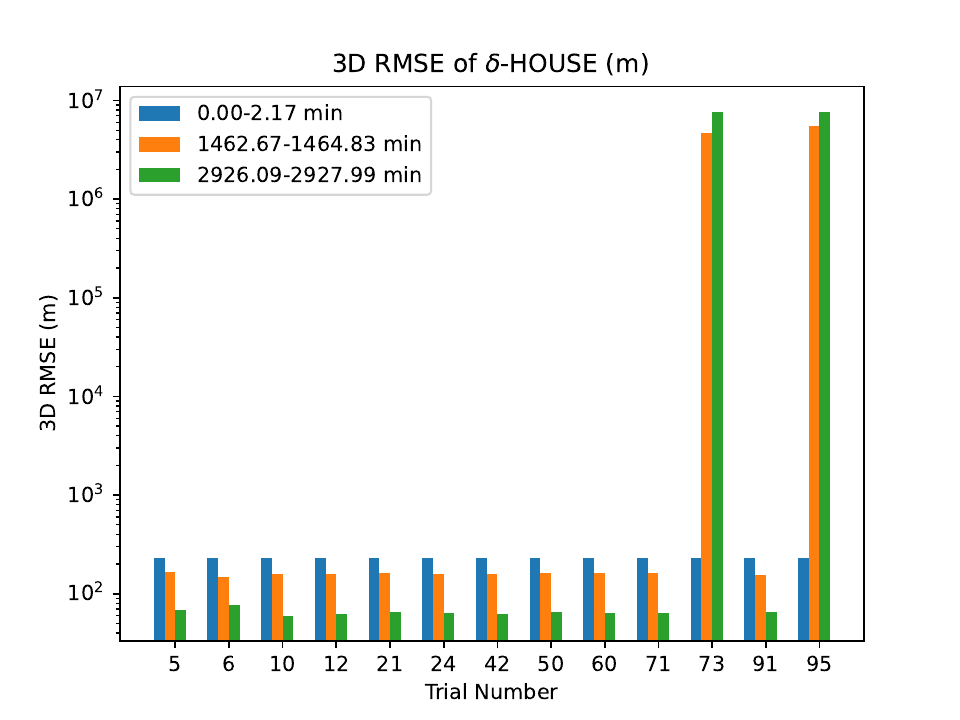}
    \caption{3D Position RMSE Values w.r.t. Various $\delta$ Values in $\delta$-HOUSE}
    \label{fig:house_3d_rmse_vs_delta}
\end{figure}

\textcolor{red}{In the case of $\delta$-HOUSE, only 13 out of 100 sets of $\delta$ values generated valid OD solutions, including 2 values leading to divergent solutions. As depicted in Fig.~\ref{fig:house_3d_rmse_vs_delta}, trial numbers 73 and 95 resulted in large 3D position RMSE values from the the second track onwards. The remaining trials yielded converged OD solutions, with the 3D position RMSE values decreasing from the first to the last track, reaching $\SI{232.525}{\meter}$, $\SI{158.96}{\meter}$, $\SI{65.089}{\meter}$ for each time window, respectively. According to Eq.~\ref{eq:upd_kappa}, when the kurtosis becomes small, it is intentially enlarged, causing $w_{2n+1}$ to be reset to $\delta$. This condition was triggered for most of the $\delta$ values in all 100 MCS. Consequently, $w_{2n+1}$ was reset to the value of $\delta$, and other weights were recalculated. However, this enlargement of kurtosis values distorts the actual state distribution, leading to filter divergence.}

\textcolor{red}{On the other hand, all selected 100 $w$ values led to the same converged OD solution at $\SI{232.564}{\meter}$, $\SI{158.734}{\meter}$, $\SI{58.834}{\meter}$ for the three tracks, respectively. This can be attributed to the fact that the weights of sigma points at each step of $w$-HOUSE do not exceed the threshold value set by the $w$ value so the reset of the last weight $w_{2n+1}$ is never triggered. Therefore, $w$-HOUSE demonstrates superior robustness and stability in terms of parameter dependency compared to $\delta$-HOUSE. In other words, it is easier for the newly developed $w$-HOUSE to select a $w$ parameter that leads to converged estimates than it is for the original $\delta$-HOUSE.}
\subsubsection{Time Consumption}
\begin{figure}[!ht]
    \centering
    \includegraphics[width=8cm]{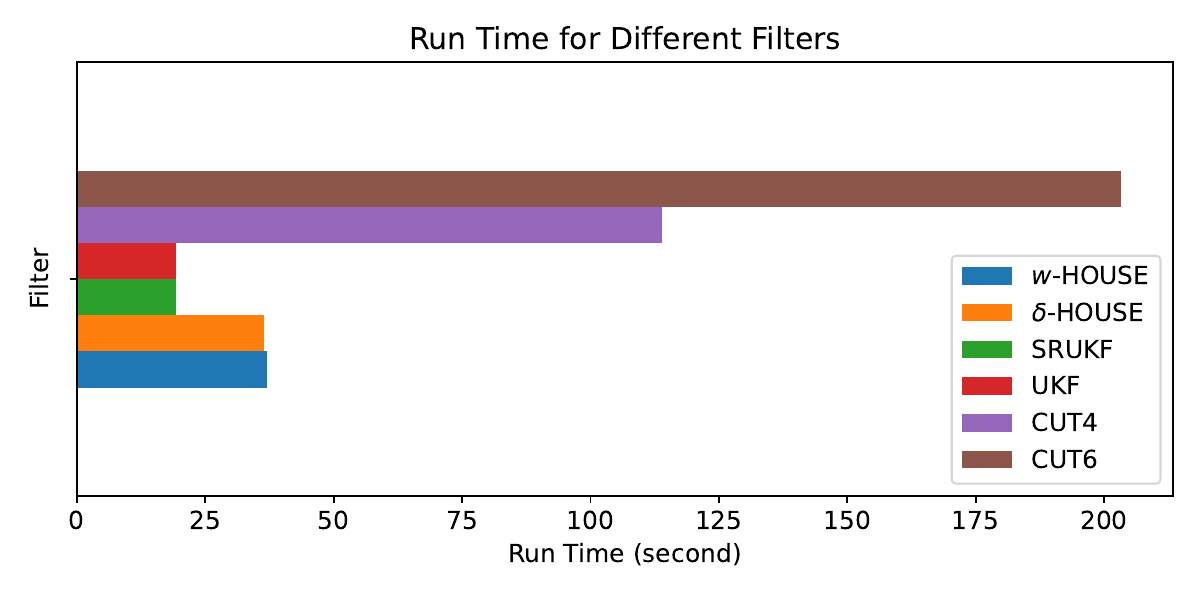}
    \caption{Runtime of Real-world OD for All Filters}
    \label{fig:all_run_time}
\end{figure}

The runtimes for all filters are shown in Fig.~\ref{fig:all_run_time}. Likewise, SRUKF and UKF are the most efficient, while CUT-4 and CUT-6 are the most time expensive. The $w$-HOUSE and $\delta$-HOUSE filters consume similar amounts of time, less than two times of that of two UKFs. The $w$-HOUSE is just slightly more time consuming than $\delta$-HOUSE.

\section{Conclusion}
\label{sec:conclusion}
This paper introduces a novel variant of the high-order unscented estimator called $w$-HOUSE. Building upon the original $\delta$-HOUSE, the author proposes the adoption of the square-root form of the covariance matrix. This modification ensures a more numerically stable evolution of the probability density function and alleviates the constraint imposed by the large kurtosis of the state distribution on the moving weight of one sigma point.

The performance of the newly proposed $w$-HOUSE was evaluated and compared with several existing filters, including square-root unscented Kalman filter, unscented Kalman filter, the conjugate unscented transformation family (CUT-4 and CUT-6), and the original $\delta$-HOUSE. The evaluation was conducted in the context of three numerical examples, including a real-world angle-only orbit determination problem of resident space objects, by considering estimation accuracy \& precision, robustness, and computational complexity. Additionally, a comparative analysis was carried out between $\delta$-HOUSE and $w$-HOUSE regarding their parametric dependency.

To facilitate orbit prediction within the OD problem, the author has developed a high-fidelity orbital propagator based on modified equinoctial elements. The scenarios utilise real-world tracking data, in the format of two angles, collected for the Sentinel 6A satellite in low Earth orbit. The processing results reveal that the proposed $w$-HOUSE method outperforms other filters in terms of positioning accuracy. \textcolor{red}{Remarkably, it achieves a 3D position RMSE of less than \SI{60}{\meter}, even with imperfect angle-only measurements contaminated with frequent outliers, in a three-day OD scenario. Specifically, the $w$-HOUSE produces a 3D position RMSE that is more than \SI{1}{\meter} smaller compared to $\delta$-HOUSE.}

Both $w$-HOUSE and $\delta$-HOUSE exhibit significantly reduced computational complexity, which is comparable to that of CUT filters. Furthermore, the paper reveals that $w$-HOUSE demonstrates greater robustness than $\delta$-HOUSE concerning the selection of the dependent parameter. Consequently, the proposed $w$-HOUSE emerges as a favourable choice for nonlinear and non-Gaussian estimation problems, particularly in vital OD. 

The availability of an OD toolbox, equipped with the aforementioned collection of derivative-free filters, will provide the astrodynamics community with valuable insights into emerging OD techniques for RSOs in LEO and other orbits.

\section{Acknowledgement}
The author would like to acknowledge Dr Zhenwei Li from Changchun Observatory of National Astronomical Observatory, Chinese Academy of Sciences for providing valuable optical space tracking measurements and discussion on the measurement quality. The preliminary implementation of the orbit determination toolbox by the undergraduate student James Appleton supported by the UNSW Engineering Taste of Research grant is also acknowledged. The author would also like to acknowledge the use of ChatGPT 3.5 for polishing the writing of this paper. 

\begin{appendices}
\section*{Appendix: Sigma Point Sampling Based Linear Least Mean Squares Estimators}
\label{app:llmse}
The process of sigma point sampling based linear least mean squares estimation involves several steps. 
The first step is the generation of samples for the augmented vector of state variables and measurement noise. Here, $N$ sigma points are generated based on the moments of the PDF of the augmented state, along with the associated weights for each sigma point. The set of sigma points and weights is represented as $\{(\vec{\boldsymbol{x}}^{(j)}(k-1),\vec{\boldsymbol{\omega}}^{(j)}(k-1),w^{(j)}):j=1,2,...N\}$, with the weights normalised such that $\sum_{j=1}^{N}w^j=1$.

Following the generation of samples, the next step is the time update. In this step, the sigma points are propagated through the nonlinear dynamics model to obtain the predicted sigma points $\vec{\boldsymbol{x}}^{(j)}(k|k-1)$. This is mathematically represented as:
\begin{gather}
    \vec{\boldsymbol{x}}^{(j)}(k|k-1) = \boldsymbol{F}\big(\vec{\boldsymbol{x}}^{(j)}(k-1),\vec{\boldsymbol{\omega}}^{(j)}(k-1)\big).
\end{gather}
The predicted mean and covariance of these sigma points are then calculated as follows:
\begin{gather}
    {\vec{\boldsymbol{x}}}_{m}(k|k-1)=\sum_{j=1}^{N} w^{(j)}\vec{\boldsymbol{x}}^{(j)}(k|k-1),     
\end{gather}
\begin{gather}
    {\boldsymbol{P}}_x(k|k-1)=\sum_{j=1}^{N} w^{(j)}\vec{\delta\boldsymbol{x}}^{(j)}(k|k-1)\Big[\vec{\delta\boldsymbol{x}}^{(j)}(k|k-1)\Big]^T.    
\end{gather}

The process then moves to the second round of sample generation for the augmented vector of state variables and measurement noise. The set of sigma points and weights is represented as $\{(\vec{\boldsymbol{x}}^{(j)}(k),\vec{\boldsymbol{\nu}}^{(j)}(k),w^{(j)}):j=1,2,...N\}$, with the weights normalised such that $\sum_{j=1}^{N}w^j=1$.

The final step is the measurement update. The deviations of the sigma points are calculated as follows:
\begin{gather}
    \vec{\delta\boldsymbol{x}}^{(j)}(k) =\vec{\boldsymbol{x}}^{(j)}(k) - {\vec{\boldsymbol{x}}}_{m}(k|k-1).  
\end{gather}
The predicted sigma points are propagated through the measurement model to obtain the predicted measurements $\vec{\boldsymbol{z}}^{(j)}(k)$:
\begin{gather}
    {\vec{\boldsymbol{z}}}^{(j)}(k) = \boldsymbol{h}\big(\vec{\boldsymbol{x}}^{(j)}(k),\vec{\boldsymbol{\nu}}^{(j)}(k)\big),
\end{gather}
followed by the calculation of the predicted mean and covariance of measurements as follows:
\begin{gather}
    {\vec{\boldsymbol{z}}}_m(k)=\sum_{j=1}^{N} w^{(j)}\vec{\boldsymbol{z}}^{(j)}(k),     
\end{gather}
\begin{gather}
    \vec{\delta\boldsymbol{z}}^{(j)}(k) =\vec{\boldsymbol{z}}^{j}(k) - {\vec{\boldsymbol{z}}}_m(k),
\end{gather}
\begin{gather}
    {\boldsymbol{P}}_{z}(k)=\sum_{j=1}^{N} w^{(j)}\vec{\delta\boldsymbol{z}}^{(j)}(k)\Big[\vec{\delta\boldsymbol{z}}^{(j)}(k)\Big]^T.
\end{gather}
Subsequently, the Kalman gain is calculated as follows:
\begin{gather}
    {\boldsymbol{K}}(k)={\boldsymbol{P}}_{x}(k|k-1){\boldsymbol{P}}_{z}^{-1}(k),
\end{gather}
based on which the state estimate is updated as follows:
\begin{gather}
    \vec{\boldsymbol{x}}(k) = \vec{\boldsymbol{x}}(k|k-1) + {\boldsymbol{K}}(k)(\vec{\boldsymbol{z}}(k) - {\vec{\boldsymbol{z}}}_m(k)).
\end{gather}

In addition to the above, other orders of moments can also be updated. For instance, the update of the second-order moment, i.e., the covariance, is also necessary by using the Kalman gain as follows:
\begin{gather}
    {\boldsymbol{P}}_{x}(k)={\boldsymbol{P}}_{x}(k-1)-{\boldsymbol{K}}(k){\boldsymbol{P}}_{z}^{-1}(k){\boldsymbol{K}}^T(k).
\end{gather}

\end{appendices}

\FloatBarrier

\bibliographystyle{IEEEtran}
\bibliography{orbDet}

\end{document}